\documentclass[aps,prd,showpacs,preprintnumbers]{revtex4}
\usepackage{amsmath}
\usepackage{subfig}
\usepackage{graphicx}
\usepackage{epstopdf}
\usepackage{float}
\usepackage{subfloat}
\usepackage{amssymb,amsfonts}
\usepackage{latexsym}
\usepackage{epsfig}
\usepackage{amssymb}
\usepackage{color}
 \usepackage{bm}
\newcommand{\eq}{\begin{equation}}
\newcommand{\eeq}{\end{equation}}
\newcommand{\eqn}{\begin{eqnarray}}
\newcommand{\eeqn}{\end{eqnarray}}
\newcommand{\non}{\nonumber\\}
\begin{document}
\title{Mass splitting of vector meson and spontaneous spin polarization under rotation}
\author{Minghua Wei$^{1,2}$ }
\thanks{weimh@mail.ihep.ac.cn, first author}
\author{Yin Jiang$^{3}$}
\thanks{jiang\_y@buaa.edu.cn, co-correspondence author}
\author{Mei Huang$^{1}$}
\thanks{huangmei@ucas.ac.cn,co-correspondence author}
\affiliation{$^{1}$ School of Nuclear Science and Technology, University of Chinese Academy of Sciences, Beijing 100049, China}
\affiliation{$^{2}$ Institute of High Energy Physics, Chinese Academy of Sciences, Beijing 100049, P.R. China}
\affiliation{$^{3}$ Department of Physics, BeiHang University, Beijing 100191, P.R. China}

\begin{abstract}
In the present paper, we study the effect of the rotation on the masses of scalar meson as well as vector meson in the framework of 2-flavor Nambu--Jona-Lasinio model. The existence of rotation causes a tedious quark propagator and corresponding polarization function. Applying the random phase approximation, the meson mass is calculated numerically. It is found that the behavior of scalar and pseudoscalar meson masses under the angular velocity $\omega$ is similar to that at finite chemical potential, both rely on the behavior of constituent quark mass and reflect the property related to the chiral symmetry. However, masses of vector meson  $\rho$ have more profound relation with rotation. After tedious calculation, it turns out that at low temperature and small chemical potenial, the mass for spin component $s_z=0,\pm 1$ of vector meson under rotation shows very simple mass splitting relation $m_{\rho}^{s_z}(\omega)=m_\rho(\omega=0)-\omega s_z$, similar to the Zeeman splitting of charged meson under magnetic fields. Especially it is noticed that the mass of spin component $s_z=1$ vector meson $\rho$ decreases linearly with $\omega$ and reaches zero at $\omega_c=m_\rho(\omega=0)$, this indicates the system will develop $s_z=1$ vector meson condensation and the system will be spontaneously spin polarized under rotation.
\end{abstract}
\pacs{12.38.Mh, 25.75.Nq, 25.75.-q }
\maketitle

\section{Introduction}
In non-central heavy-ion collision(HIC), large vorticity and strong magnetic field are expected to be generated in extremely hot quark gluon plasma(QGP). Straightforward electromagnetical(EM) computation shows the magnetic field would reach about $O(10^{14})T$\cite{Kharzeev:2007jp}
in the early stage of HIC, while kinetic and hydrodynamic simulations\cite{Becattini:2007sr,Jiang:2016woz} indicate the  local vorticity would exceed $0.5 fm^{-1}$ with the total angular momentum of QGP at a range of $O(10^{4})-O(10^{5})\hbar$. Known as the Barnett and magnetization effects, spin particles are polarized by these pseudo vector field and thus distribute differently from the normal thermal distributions. Besides chiral effects induced by such pseudo vector fields\cite{Kharzeev:2007tn, Son:2009tf, Kharzeev:2010gr}, studies on these distribution modifications would be helpful to understand the hadronization mechanism of the strong interaction as well. Inspired by the large amplitude and retention by the angular momentum conservation, vorticity has attracted more and more interests recently.

%Show the experimental results involving with the magnetic field and especially vorticity.
Comparing with magnetic field effects, the rotation-related effects are electric charge blind, and only involve kinetic properties of the QGP and  strong interaction which we are mostly interested in. Experimentally, in order to screen out the EM effects neutral particles with finite spin numbers are chosen as carriers of the vorticity polarization effects. As it is difficult to detect the chargeless particle directly the distribution of its charged daughter particle serves as an alternative observable for the global polarization effect. With the help of the $\Lambda$ hyperon the average magnitude of the vorticity of QGP has been extracted by the STAR collaboration\cite{STAR:2017ckg}. In these measurements the expectation of $\Lambda$ polarization as well as the vorticity behavior of collision energy have been confirmed as well. All the results seem to be understandable by considering the energy shift induced by the voriticity polarization to spins. However the theory became a little vague when the $K^{*0}$ and $\phi$ mesons' measurements were presented in \cite{Acharya:2019vpe}. The mismatch between these measurement indicates the fine structure of hadrons may play an non-negligible role in polarization processes.

The mass is one of the most fundamental attributes of a hadron. For a composite particle it will be modified by the single-particle dispersion relation of the fundamental degree of freedom as well as the interaction among them. Studies of hadron masses would help us to discover many clues of the environment where hadrons are born. As a well-known example, $\sigma$ meson and pion masses would change with the growing temperature and chemical potential because of chiral restoration~\cite{Klevansky:1992qe}. And recently people have studied the vector meson $\rho$ mass in external magnetic field as well by taking the polarization effect on quarks into account. A lattice calculation demonstrates that charged $\rho$ meson mass decreases firstly and increases finally, leaving a minimum around $eB\simeq 1GeV^{2}$~\cite{Hidaka:2012mz}. And by using effective models, such as the Nambu--Jona-Lasinio(NJL) model with vector channel, the $\rho$ meson with different spin components have been studied~\cite{Liu:2014uwa,Liu:2018zag}. Therefore a natural question is what about the mass behavior under a background vorticity field which is a little like the magnetic case at the first sight. For the rotating effect, the co-rotating frame~\cite{Yamamoto:2013zwa} is usually adopted and a nontrivial spin connection term will be introduced~\cite{Matsuo:2015}, which serves as a polarization term for angular momentums. With this extended NJL model it is suggested that chiral phase transition would take place as angular velocity increasing~\cite{Jiang:2016wvv}. Furthermore, people have established more complicated phase diagrams which combine rotation and other physical conditions such as chemical potential, isospin and magnetic field~\cite{Wang:2018sur,Zhang:2018ome,Chen:2015hfc}. In those NJL models, at the quark level, the rotation always behaves as an effective chemical potential. This analogy has been understood with a  Hamiltonian shifting $\hat{H}\rightarrow\hat{H}-\vec{\omega}\cdot\hat{J}$ and the latter term may be corresponding to an effective chemical potential~\cite{Chen:2015hfc,Matsuo:2012wv}. At the same quark level, holographic models also contribute to elaborate the property of rotating quark matter by setting up a four-dimensional AdS-Kerr-Newman black hole to construct a rotation-magnetism analogy~\cite{McInnes:2016dwk}. While for the composite hadrons, such as vector mesons, there have been few works on the mass behaviors.

%The mass of $\rho^{\pm}$ with $S_{z}=0$ will increase as the magnetic field growing, while the mass of $\rho^{+}(\rho^{-})$ with $S_{z}=+1(S_{z}=-1)$ will decrease linearly with $eB$.
%Most of these data could be understood by noticing the energy shift induced by the polarization effects on partons and further the consequent hadrons.
%The intriguing properties of QCD matter under those extreme conditions appeal people to investigate them with theoretical methods.

%The  motivation of this paper is calculating masses of vector mesons, i.e. the XXX, in the two-flavor NJL model. The quark propagator in the rotating frame is similar to the quark propagator under the external magnetic field~\cite{Miransky:2015ava}. The Laguerre polynomial will vanish instead of Bessel function, and the summation of Landau level corresponding to the summation of all the possible order of Bessel function. However, the final numerical result still sustain the analogy of rotation and chemical potential.

In this paper, we focus on the scalar and the vector meson and investigate their masses under the rotation at finite chemical potential. In Sec. ~\ref{NJL model under rotating frame}, in order to deal with both the finite temperature and density cases, we introduce the two- flavor NJL model with vector channel in the co-rotating frame. In this framework, we generate the dynamical quark mass with chiral symmetry spontaneous breaking and construct scalar and vector masons with the dressed quark propagator and extract the corresponding masses with the well-known random phase approximation(RPA) in Sec. ~\ref{Scalar and vector meson mass under rotating frame} and show their numerical results in Sec. ~\ref{Numerical results and discussion}. Because of the rich phase structure at large chemical potential we only study the  range of $\mu_q<200$MeV in this work and leave the discussion of the rotating color superconductivity in our following works.  We have found that masses of scalar mesons are controlled by the chiral phase transition which could be driven by temperature, density and rotation. While the vector meson, which carries net angular momentum, is governed by the polarization effect on the total angular momentum before chiral symmetry restoration. At large angular velocity the  mass of spin component $s_z=1$ for the vector meson vanishes. This indicates the macroscopic condensate of spin component $s_z=1$ of vector meson $\langle \rho^{s_z=1}\rangle$  thus spontaneous spin polarization would be induced in the ultra-fast rotating system. In Sec.~\ref{Conclusion},  we summarize our main results and give an outlook.

\section{NJL model in co-rotating frame}
\label{NJL model under rotating frame}
NJL model is an effective model with 4-fermion interaction which is widely used to study quark-quark and quark-antiquark pairing which corresponding to chiral phase transition, superfluidity and superconductivity and so on. Besides the usual scalar channels we take account the vector channels in order to construct the vector $\rho$ mesons. The Lagrangian of the two-flavor NJL model in the co-rotating frame is given by~\cite{Wang:2018sur,Bernard:1988db}:
\eq
\mathcal{L} = \bar{\psi}[i\bar{\gamma}^{\mu}(\partial_{\mu}+\Gamma_{\mu})-m]\psi+G_S[(\bar{\psi}\psi)^2+(\bar{\psi}i\gamma_5\vec{\tau}\psi)^2]-G_V[(\bar{\psi}\gamma_{\mu}\psi)^2+(\bar{\psi}\gamma_{\mu}\gamma_5\psi)^2],
\eeq
where $m$ is the current quark mass. $G_{S}$ and $G_{V}$ are the coupling constants in the scalar and vector channels, respectively. In the curved co-rotating frame the gamma matrices $\bar{\gamma}^{\mu}$ should be defined according to the corresponding Clifford algebra. The curved gamma matrices are connected with the flat ones with the vierbein as $\bar{\gamma}^{\mu}=e_{a}^{\ \mu}\gamma^{a}$ and where $e_{a}^{\ \mu}$ should be chosen to satisfy $g_{\mu\nu}=\eta_{ab}e^{a}_{\ \mu}e^{b}_{\ \nu}$, where $\eta_{ab}$ is the metric of flat space-time and $\gamma^{a}$ is flat gamma matrices. In our case a simple enough choice is $e^{a}_{\ \mu}=\delta^a_{\ \mu}+  \delta^a_{\ i}\delta^0_{\ \mu} \, v_i$ and $e_{a}^{\ \mu}=\delta_a^{\ \mu} -  \delta_a^{\ 0}\delta_i^{\ \mu} \, v_i$, where $v_i$ is the linear velocity $\vec{v} =\vec{\omega}\times\vec{x}$ under the presence of  a constant angular velocity $\vec{\omega}$. The so-called spinor connection is given by $\Gamma_\mu=\frac{1}{4}\times\frac{1}{2}[\gamma^a,\gamma ^b] \, \Gamma_{ab\mu}$, where $\Gamma_{ab\mu}=\eta_{ac}(e^c_{\ \sigma} G^\sigma_{\ \mu\nu}e_b^{\ \nu}-e_b^{\ \nu}\partial_\mu e^c_{\ \nu})$ and $G^\sigma_{\ \mu\nu}$ is the usual Christoffel connection determined by $g_{\mu\nu}$~\cite{Yamamoto:2013zwa,Jiang:2016wvv,Matsuo:2015}. In the slow velocity limit $|\vec{\omega}\times\vec{x}|\ll c$ we could only keep the $O(\vec{v})$ terms which can be reduced to the ordinary polarization form as $\vec{\omega}\cdot\vec{J}$, where $\vec{J}=\vec{x}\times\vec{p}+\vec{S}$ is the total angular momentum~\cite{Jiang:2016wvv,Matsuo:2015} and $\vec{S}=\frac{1}{2}\left(
                      \begin{array}{cc}
                        \vec{\sigma} & 0 \\
                        0 & \vec{\sigma} \\
                      \end{array}
                    \right)
$ is the spin operator.

Applying the mean field approximation and choosing the direction of rotation as the $z$-axis, the bilinear part of the Lagrangian at finite chemical potential is given by~\cite{Wang:2018sur}
\eq
\mathcal{L}=\bar{\psi}[i\gamma^{\mu}\partial_{\mu}+\gamma^{0}(\omega \hat{J_{z}}+\mu)-M]\psi-\frac{(M-m)^{2}}{4G_{S}},
\eeq
where $J_{z}$ is the third component of total angular momentum $\vec{J}$, and $\mu$ is the quark chemical potential, it is seen that the angular velocity plays similar role as the chemical potential, and $M$ is the constituent quark mass which is given by the chiral condensate as $M = m - 2G_S\left<\bar\psi\psi\right>$.
The general grand potential is given by~\cite{Jiang:2016wvv,Wang:2018sur}:
\eqn
\label{omg}
\Omega(T,\mu,M,\omega) &=&\int d^3 \mathbf{r}~ \bigg\{\frac{(M-m)^2}{4G_S} \nonumber \\
     & & -\frac{N_c N_f}{16\pi^2} T \sum_n\int dk_t^2 \int dk_z[J_n(k_t r)^2+J_{n+1}(k_t r)^2]\left[ \ln(1+e^{(E_k-(n+\frac{1}{2})\omega-\mu)/T})\right.\non
& & + \left. \ln(1+e^{-(E_k-(n+\frac{1}{2})\omega-\mu)/T})+\ln(1+e^{-(E_k+(n+\frac{1}{2})\omega+\mu)/T})+ \ln(1+ e^{(E_k+(n+\frac{1}{2})\omega+\mu)/T})\right]\bigg\}.\non
\eeqn
where $E_k=\sqrt{k_t^2+k_z^2+M^2}$ and $k_{t, z}$ are the transverse and longitudinal momentum respectively. Obviously the local potential approximation $\partial_r M(r)\simeq 0$ has been adopted during solving the eigen modes. In the following computation we choose $N_c=3$ and $N_f=2$. In this work we will neglect the four-fermion contributions to the ground state, which means the chiral condensate is completely computed by the gap equation as
$\frac{\partial \Omega}{\partial M} = 0$
with the constraint $\frac{\partial^2 \Omega}{\partial M^2} > 0$. In Sec. \ref{Numerical results and discussion}, we will show the numerical result of the constituent quark mass $M$. It serves as the environment where mesons are given birth, and thus modifies their masses. In the mean field approximation the gap equation is just the one-loop diagram of the quark propagator which reads as
\eq
\label{propagator}
\begin{aligned}
S(\tilde{r};\tilde{r'})&=\frac{1}{(2 \pi )^2}\sum _n\int\frac{d k_0}{2\pi } \int k_t d k_t \int d k_z \frac{e^{i n \left(\phi -\phi '\right)}e^{-i k_0 \left(t-t'\right)+i k_z \left(z-z'\right)}}{[k_{0}+(n+\frac{1}{2})\omega]^{2}-k_{t}^{2}-k_{z}^{2}-M^{2}+i\epsilon} \\
&
\times\{[[k_{0}+(n+\frac{1}{2})\omega]\gamma^{0}-k_{z}\gamma^{3}+M][J_{n}(k_{t}r)J_{n}(k_{t}r')\mathcal{P}_{+}+e^{i (\phi-\phi)'}J_{n+1}(k_{t}r)J_{n+1}(k_{t}r')\mathcal{P}_{-}]\\
&-i\gamma^{1}k_{t}e^{i \phi}J_{n+1}(k_{t}r)J_{n}(k_{t}r')\mathcal{P}_{+}-\gamma^{2}k_{t}e^{-i\phi '}J_{n}(k_{t}r)J_{n+1}(k_{t}r')\mathcal{P}_{-}
\},
\end{aligned}
\eeq
where $\mathcal{P}_{\pm}=\frac{1}{2}(1\pm i\gamma^{1}\gamma^{2})$ are projection operators and $\tilde{r}=(t,r,\theta,\phi)$ are the coordinates in the cylindrical frame.

\section{Scalar and vector meson mass under rotation}
\label{Scalar and vector meson mass under rotating frame}
\subsection{The scalar meson}
In the NJL model, meson is regarded as  $q \bar q $ bound states or resonances, which can be obtained from the quark-antiquark scattering amplitude
\cite{Buballa:2003qv,He:1997gn,Rehberg:1995nr}. In the random phase approximation (RPA), the full propagator of $\sigma$ meson $D_{\sigma}(q^2)$ can be expressed to leading order in $1/N_c$ as an infinite sum of quark-loop chains:
\begin{equation}\label{meson propagator}
	D_{\sigma}(q^2)=\frac{2G_{S}}{1-2G_{S}\Pi_{s}(q^2)},
\end{equation}
where $\Pi_{s}(q^2)$ is the quark one-loop polarization function and takes the form of
\begin{eqnarray}
\label{pola}
\Pi_{s}(q)=-i \int d^{4}\tilde{r}Tr_{sfc}[i S(0;\tilde{r})i S(\tilde{r};0)]e^{i q\cdot \tilde{r}},
\end{eqnarray}
 where $Tr_{sfc}$ means trace in spin, flavor and color space.  After a tedious calculation in Appendix.~\ref{appendix:a}, the polarization function could be simplified as this form
\eq
\begin{aligned}
\Pi_{s}(q^2)&=
-2 i N_{f}N_{c}\int \frac{d^{4} p}{(2\pi)^{4}} \\
&\times
\left\{ \frac{\left(p_{0}+q_{0}+\frac{1}{2}\omega \right) \left(p_{0}+\frac{1}{2}\omega \right)+M^2-(\vec{p}+\vec{q})\cdot \vec{p}}
{\left[\left(p_{0}+q_{0}+\frac{1}{2}\omega\right)^{2}-(\vec{p}+\vec{q})^{2}-M^{2}\right]
\left[\left(p_{0}+\frac{1}{2}\omega\right)^{2}-\vec{p}^{2}-M^{2}\right]}\right.\\
&+ \left.\frac{\left(p_{0}+q_{0}-\frac{1}{2}\omega \right) \left(p_{0}-\frac{1}{2}\omega \right)+M^2-(\vec{p}+\vec{q})\cdot \vec{p}}
{\left[\left(p_{0}+q_{0}-\frac{1}{2}\omega\right)^{2}-(\vec{p}+\vec{q})^{2}-M^{2}\right]
\left[\left(p_{0}-\frac{1}{2}\omega\right)^{2}-\vec{p}^{2}-M^{2}\right]}
\right\}.
\end{aligned}
\eeq
If we use finite temperature theory with chemical potential \cite{Kapusta}, the polarization function will be:
\begin{equation}
	\Pi_{s}(\vec{q},i\nu_{n})=2N_{f} N_{c}T \sum_{s=\pm}\sum_N\int \frac{d^{3}\vec{p}}{(2\pi)^{3}}
	\frac{[(i \tilde{\omega}_{N}+i \nu_{n})+\frac{1}{2}s\omega+\mu][i \tilde{\omega}_{N}+\frac{1}{2}s\omega+\mu]+M^{2}-(\vec{p}+\vec{q})\cdot\vec{p}}{[(i \tilde{\omega}_{N}+i \nu_{n}+\frac{1}{2}s\omega+\mu)^{2}-(\vec{p}+\vec{q})^{2}-M^{2}][(i \tilde{\omega}_{N}+\frac{1}{2}s\omega+\mu)^{2}-\vec{p}^{2}-M^{2}]},
\end{equation}
where $\tilde{\omega}_{N}=(2N+1)\pi T$ is  Matsubara frequency.
Considered analytic continuation $\Pi_{s}(\vec{q},\tilde{\nu})=\Pi_{s}(\vec{q},i\nu_{n})|_{\tilde{\nu}+i \eta}$ and set $\vec{q}=0$, an explicit form of $\Pi_{s}(0,\tilde{\nu})$ is shown in Appendix~\ref{appendix:a}

From the pole of above propagator in Eq.(\ref{meson propagator}), the $\sigma$ mass can be obtained by solving:
\begin{equation}
	1-2 G_{S} \Pi_{s}(0,\tilde{\nu})=0,
\end{equation}
We have similar operation for pseudoscalar meson $\pi$. The operators in polarization functions are defined as $\tau^{\pm}=\frac{1}{\sqrt{2}}(\tau_{1}\pm i \tau_{2})$ where $\tau_{i}$ are Pauli Matrice. In polarization functions, we choose $\tau^a=\tau^3,\tau^b=\tau^3$ for neutral pion and $\tau^a=\tau^+,\tau^b=\tau^-$ for charged pion. However, polarization functions have the same form for different charged mesones. 
\eq
\begin{aligned}
\Pi_{ps}(q^2)&=-i \int d^{4}\tilde{r}Tr_{sfc}[i \gamma^{5}\tau^{a}i S(0;\tilde{r})i \gamma^{5}\tau^{b}i S(\tilde{r};0)]e^{i q\cdot \tilde{r}}\\
&=4 i N_{f}N_{c}\int \frac{d^{4} p}{(2\pi)^{4}} \\
&\times
\left\{ \frac{\left(p_{0}+q_{0}+\frac{1}{2}\omega \right) \left(p_{0}+\frac{1}{2}\omega \right)-M^2-(\vec{p}+\vec{q})\cdot \vec{p}}
{\left[\left(p_{0}+q_{0}+\frac{1}{2}\omega\right)^{2}-(\vec{p}+\vec{q})^{2}-M^{2}\right]
\left[\left(p_{0}+\frac{1}{2}\omega\right)^{2}-\vec{p}^{2}-M^{2}\right]}\right.\\
&+ \left.\frac{\left(p_{0}+q_{0}-\frac{1}{2}\omega \right) \left(p_{0}-\frac{1}{2}\omega \right)-M^2-(\vec{p}+\vec{q})\cdot \vec{p}}
{\left[\left(p_{0}+q_{0}-\frac{1}{2}\omega\right)^{2}-(\vec{p}+\vec{q})^{2}-M^{2}\right]
\left[\left(p_{0}-\frac{1}{2}\omega\right)^{2}-\vec{p}^{2}-M^{2}\right]}.
\right\}
\end{aligned}
\eeq
For finite temperature formalism with chemical potential, the polarization function will be:
\begin{equation}
	\Pi_{ps}(\vec{q},i\nu_{n})=-4N_{f} N_{c}T \sum_{s=\pm}\sum_N\int \frac{d^{3}\vec{p}}{(2\pi)^{3}}
	\frac{[(i \tilde{\omega}_{N}+i \nu_{n})+\frac{1}{2}s\omega+\mu][i \tilde{\omega}_{N}+\frac{1}{2}s\omega+\mu]-M^{2}-(\vec{p}+\vec{q})\cdot\vec{p}}{[(i \tilde{\omega}_{N}+i \nu_{n}+\frac{1}{2}s\omega+\mu)^{2}-(\vec{p}+\vec{q})^{2}-M^{2}][(i \tilde{\omega}_{N}+\frac{1}{2}s\omega+\mu)^{2}-\vec{p}^{2}-M^{2}]}.
\end{equation}
Considered analytic continuation $\Pi_{ps}(\vec{q},\tilde{\nu})=\Pi_{ps}(\vec{q},i\nu_{n})|_{\tilde{\nu}+i \eta}$ and set $\vec{q}=0$, an explicit form of $\Pi_{ps}(0,\tilde{\nu})$ is shown in Appendix~\ref{appendix:a}

From the pole of above propagator, the pion mass can be obtained by solving:
\begin{equation}
	1-2 G_{S} \Pi_{ps}(0,\tilde{\nu})=0.
\end{equation}

\subsection{The $\rho$ meson}
Following the Ref.~\cite{Liu:2014uwa}, we construct the vector meson in a similar way with the rotation-modified quark propagators. For the 2-flavor model we take the vector $\rho$ meson for example, its 1-loop polarization function reads as
\eq
\Pi^{\mu\nu,ab}(q)=-i \int d^{4}\tilde{r}Tr_{sfc}[i \gamma^{\mu}\tau^{a}S(0;\tilde{r})i \gamma^{\nu}\tau^{b}S(\tilde{r};0)]e^{i q\cdot \tilde{r}}.
\eeq
As there is no isospin breaking in the quark propagators $S(0;\tilde{r})$, the polarization functions of charged and neutral $\rho$ mesons are supposed to be the same under rotation. Nonzero elements of the matrix reads as
\eq
\Pi^{\mu\nu}_{\rho}=\left(
                      \begin{array}{cccc}
                        0 & 0 & 0 & 0 \\
                        0 & \Pi^{11} & \Pi^{12} & 0 \\
                        0 & \Pi^{21} & \Pi^{22} & 0 \\
                        0 & 0 & 0 & \Pi^{33} \\
                      \end{array}
                    \right).
\eeq
The explicit expressions of matrix elements are shown in Appendix~\ref{appendix:b}. The analysis of the Lorentz structure suggests the tensor can be decomposed according to its polarization directions as follows
\eq
\Pi^{\mu\nu}_{\rho}=A_{1}^{2}P^{\mu\nu}_{1}+A_{2}^{2}P^{\mu\nu}_{2}+A_{3}^{2}L^{\mu\nu}+A_{4}^{2}u^{\mu}u^{\nu},
\eeq
where $u^{\mu}$ is the four momentum in the rest frame. $u^\mu=(1,0,0,0)$ is a unit vector. And the projection operators are given as:
\eq
\begin{aligned}
P^{\mu\nu}_{1}&=-\epsilon^{\mu}_{1}\epsilon^{\nu}_{1},(s_{z}=-1 \text{ for } \rho \text{ meson }),\\
P^{\mu\nu}_{2}&=-\epsilon^{\mu}_{2}\epsilon^{\nu}_{2},(s_{z}=+1\text{ for } \rho \text{ meson }),\\
L^{\mu\nu}&=-b^{\mu}b^{\nu},(s_{z}=0 \text{ for } \rho \text{ meson }).
\end{aligned}
\eeq
where in flat frame $\epsilon^{\mu}_{1}=\frac{1}{\sqrt{2}}(0,1,i,0)$ and $\epsilon^{\mu}_{2}=\frac{1}{\sqrt{2}}(0,1,-i,0)$ are the right and left-hand polarization vectors respectively. And $b^{\mu}=(0,0,0,1)$ is the direction of rotation. As a result the $\rho$ meson propagator can be decomposed in the similar way as:
\eq
D^{\mu\nu}_{\rho}(q^{2})=D_{1}(q^{2})P^{\mu\nu}_{1}+D_{2}(q^{2})P^{\mu\nu}_{2}+D_{3}(q^{2})L^{\mu\nu}+D_{4}(q^{2})u^{\mu}u^{\nu},
\eeq
where coefficients $D_{i}$ have the RPA summation forms as:
\eq
D_{i}(q^2)=\frac{2G_{V}}{1+2G_{V}A_{i}^2}.
\eeq
Again the momentum poles here are corresponding to masses of vector $\rho$ mesons which are solutions to equations:
\eq
\label{vectorpole}
1+2G_{V}A_{i}^{2}=0,
\eeq
where
\eq
\label{coeff}
\begin{aligned}
A_{1}^{2}&=-(\Pi_{11} - i \Pi_{12}),(s_{z}=-1 \text{ for } \rho \text{ meson }),\\
A_{2}^{2}&=-\Pi_{11} - i \Pi_{12},(s_{z}=+1\text{ for } \rho \text{ meson }),\\
A_{3}^{2}&=-\Pi_{33},(s_{z}=0 \text{ for } \rho \text{ meson }).
\end{aligned}
\eeq
%\begin{figure}[t!]
% 	\includegraphics[width=250pt]{quarkmassasfunctionofomega.pdf}\\
% 	\caption{quark mass as a function of angular velocity at different chemical potential at temperature $T=10 MeV$.} \label{fig:quarkmass}
% \end{figure}

\section{Numerical results and discussion}
\label{Numerical results and discussion}
In order to evaluate the mass of $\rho$ meson at finite chemical potential and relatively large vorticity, we choose the soft cut-off scheme to avoid the leakage of the energy scale. The cut-off function is:

\eq
f_{\Lambda}(\bm{p})=\frac{\Lambda^{10}}{\Lambda^{10}+\bm{p}^{10}},
\eeq
where $\Lambda=582$MeV. In numerical calculation, Momentum integrals are understood as follows~\cite{Frasca:2011zn}
\eq
\int \frac{d \bm{p}}{2 \pi}\rightarrow\int \frac{d \bm{p}}{2 \pi}f_{\Lambda}(\bm{p}).
\eeq
The other parameters are chosen as those in Ref~\cite{Liu:2014uwa} , i.e. $G_{S}\Lambda^{2}=2.388$ and $G_{V}\Lambda^{2}=1.73$ and the current quark mass $m_{0}=5$MeV.

By neglecting mesons' fluctuations it is easy to solve the gap equation of chiral condensate at finite temperature as well as chemical potential under rotation. As the phase diagram shown in  Ref.~\cite{Jiang:2016wvv,Wang:2018sur,Chen:2015hfc} the vorticity serves as another kind of chemical potential which would weaken the chiral condensate at finite temperature case and complement the chemical potential at finite density case. As shown in Fig.(\ref{subfig:mesonmassT3}) there is a crossover at medium temperature along the angular velocity. While at low temperature the increase of chemical potential will change the 1st order chiral restoration to a crossover in Fig.(\ref{subfig:mesonmass1}), (\ref{subfig:mesonmass2}) and (\ref{subfig:mesonmass3}). As the phase structure determines the macroscopic properties of the system it is reasonable to expect that the dependence of meson masses on the angular velocity would be smooth at medium temperature and density systems, while kinked at the 1st order point for the low density systems.
%\subsection{The constituent quark mass}
%In order to calculate the meson mass, the constituent quark mass should be calculated first. In this paper, we concentrate on the situation at a fixed temperature $T=10 MeV$, and calculate constituent quark mass as a function of angular velocity at different chemical potential at temperature $T=10 MeV$. Following Ref.~\cite{Jiang:2016wvv} we choose a particular value of transverse radial coordinate $r = 0.1 GeV^{?1}$ in our calculation. In typical Au-Au collision at RHIC energy, the QCD matter has the local angular velocity at a range of 0.01-0.1 $GeV$~\cite{Becattini:2007sr}. However, we will set the angular velocity at a range of $0-1.6 GeV$ so that chiral phase transition can be find as $\omega$ increasing.

%In Fig.~\ref{fig:quarkmass},it is shown that constituent quark mass changes as a function of angular velocity at different chemical potential at temperature $T=10 MeV$. In case of pure rotation, a first order phase transition will occur at $\omega=0.781GeV$. However, in the case of $\mu=0.1,0.2,0.3GeV$, crossover will take place as $\omega$ increasing. For $\mu=0.4GeV$, the first order phase transition reappear. This coaction of "effective chemical potential" and real chemical potential also has been investigated in previous work~\cite{Wang:2018sur}.

\subsection{The scalar meson}
\begin{figure}[t!]
\subfloat[scalar meson mass as a function of angular velocity at $\mu=0 MeV$]{\includegraphics[width=200pt]{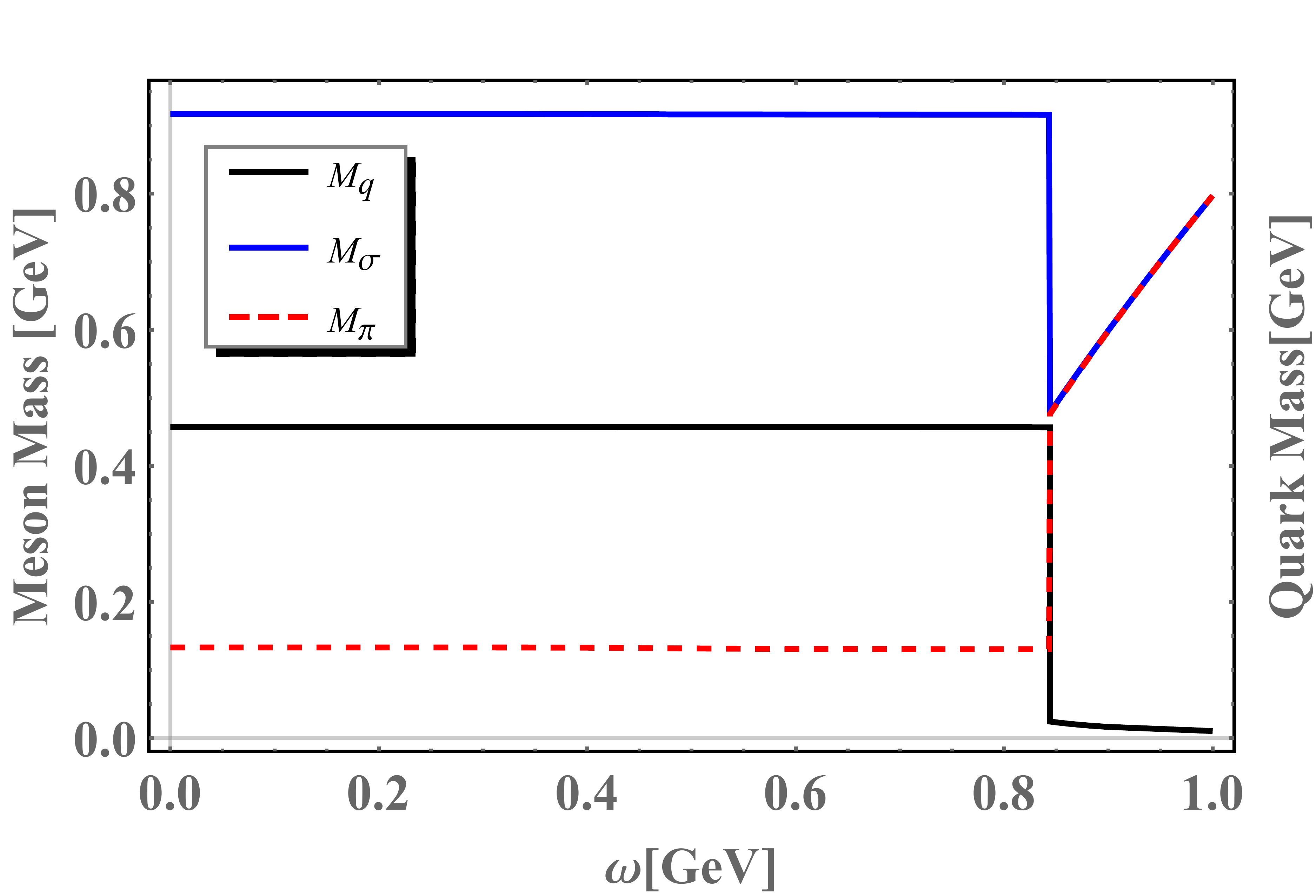}\label{subfig:mesonmass1}\hspace{1pt}}\hspace{30pt}
\subfloat[scalar meson mass as a function of angular velocity at $T=150 MeV$]{\includegraphics[width=200pt]{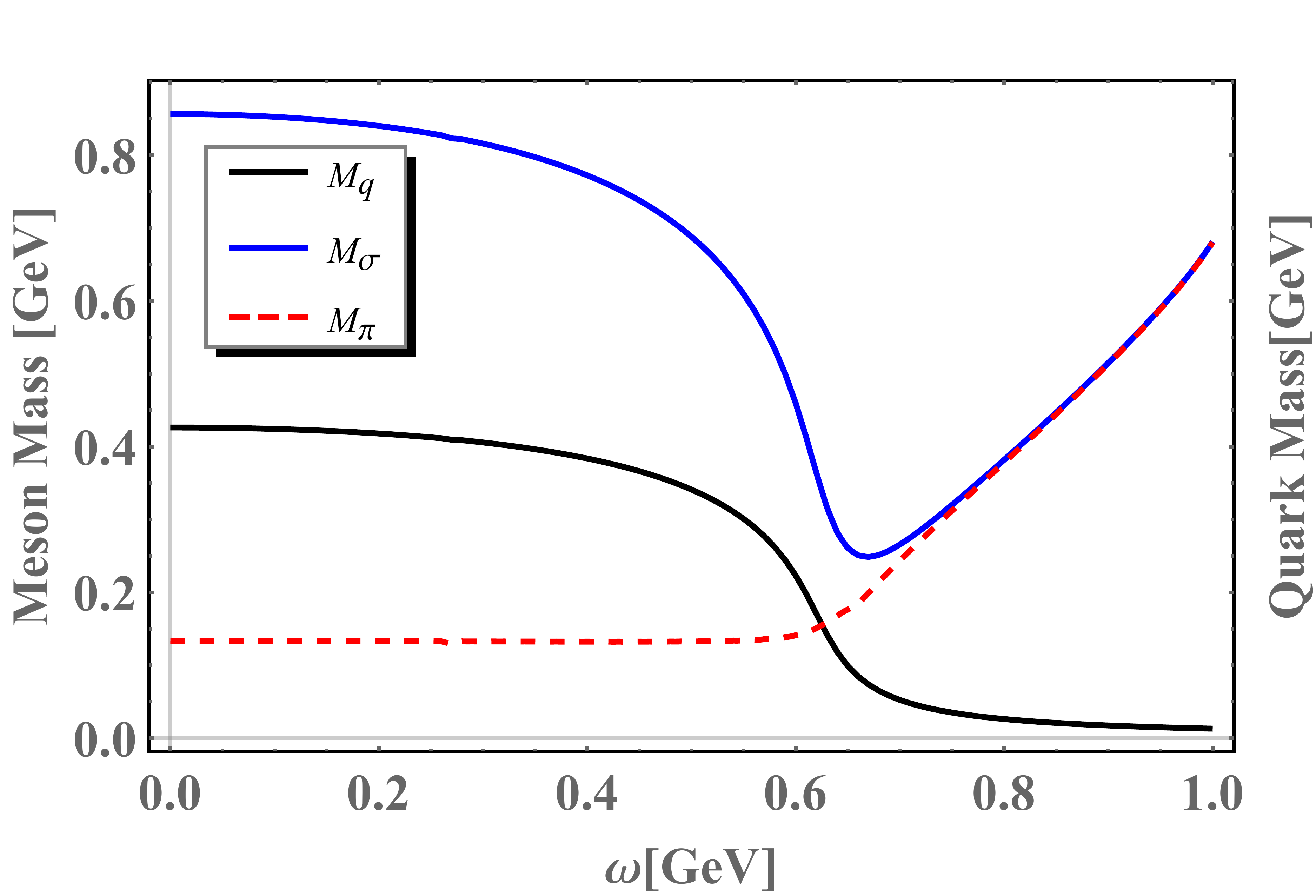}\label{subfig:mesonmassT3}\hspace{1pt}}\\
\subfloat[scalar meson mass as a function of angular velocity at $\mu=100 MeV$]{\includegraphics[width=200pt]{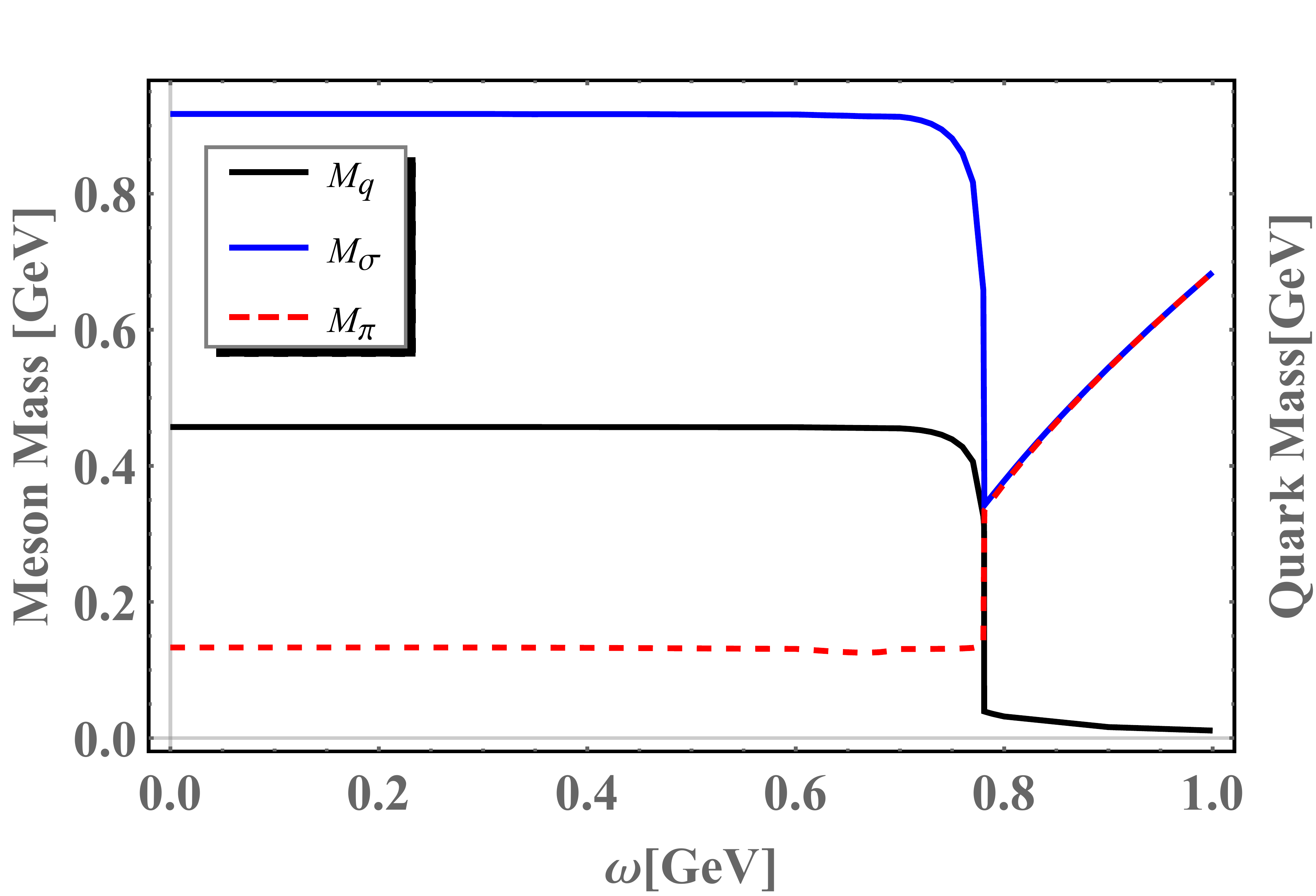}\label{subfig:mesonmass2}\hspace{1pt}}\hspace{30pt}
\subfloat[scalar meson mass as a function of angular velocity at $\mu=200 MeV$]{\includegraphics[width=200pt]{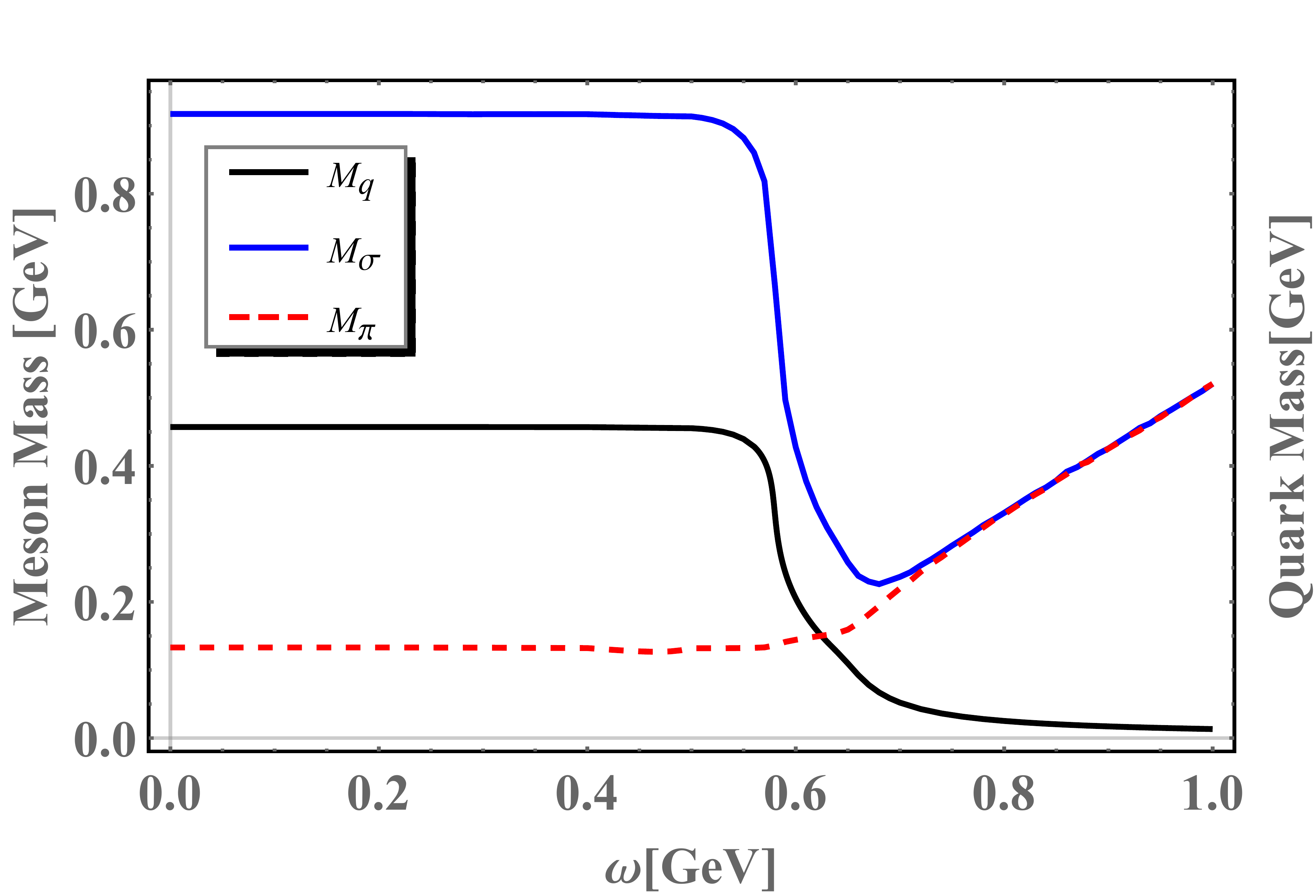}\label{subfig:mesonmass3}\hspace{1pt}}
%  \subfloat[scalar meson mass as a function of angular velocity at $\mu=300 MeV$]{\includegraphics[width=200pt]{scalarmesonmassatmu300MeV.pdf}\label{subfig:mesonmass4}\hspace{1pt}}\\
%   \subfloat[scalar meson mass as a function of angular velocity at $\mu=400 MeV$]{\includegraphics[width=200pt]{scalarmesonmassatmu400MeV.pdf}\label{subfig:mesonmass5}\hspace{1pt}}\\
\caption{scalar meson mass as a function of angular velocity at different chemical potential and temperature.}\label{fig:mesonmass}
 \end{figure}

%\begin{figure}[t!]
% 	\subfloat[scalar meson mass as a function of angular velocity at $T=50 MeV$]{\includegraphics[width=200pt]{scalarmesonmassatT50MeV.pdf}\label{subfig:mesonmassT1}\hspace{1pt}}\hspace{30pt}
% 	\subfloat[scalar meson mass as a function of angular velocity at $T=100 MeV$]{\includegraphics[width=200pt]{scalarmesonmassatT50MeV.pdf}\label{subfig:mesonmassT2}\hspace{1pt}}\\
% \subfloat[scalar meson mass as a function of angular velocity at $T=150 MeV$]{\includegraphics[width=200pt]{scalarmesonmassatT150MeV.pdf}\label{subfig:mesonmassT3}\hspace{1pt}}
%  \subfloat[scalar meson mass as a function of angular velocity at $\mu=300 MeV$]{\includegraphics[width=200pt]{scalarmesonmassatmu300MeV.pdf}\label{subfig:mesonmass4}\hspace{1pt}}\\
%   \subfloat[scalar meson mass as a function of angular velocity at $\mu=400 MeV$]{\includegraphics[width=200pt]{scalarmesonmassatmu400MeV.pdf}\label{subfig:mesonmass5}\hspace{1pt}}\\
%	\caption{scalar meson mass as a function of angular velocity at different temperature at chemical potential $\mu=0 MeV$.}\label{fig:mesonmassTemperature}
% \end{figure}
 Because of carrying no net angular momentum, the profile of scalar meson mass is completely determined by the chiral symmetry in our model. For the zero chemical potential case shown in Fig.(\ref{subfig:mesonmass1}) and (\ref{subfig:mesonmassT3}), as angular velocity increases the chiral condensate behaves the same as that in the \cite{Jiang:2016wvv}. At extremely low temperature the chiral restoration is 1st order and thus the masses keep invariant and then jump together at the critical angular velocity. While in hot matter the condensate keeps melting slowly until the crossover range $\omega\sim 0.6$GeV. As the consequence, $\sigma$ meson mass stays almost static and pions serve as Goldstone particles in the chiral breaking phase. When the $\omega$ close to the crossover range they approach each other and eventually become almost degenerate because of the chiral symmetry restoration. The behavior at finite density could be understood with chiral symmetry as well by noticing the order of phase transition. As Fig.(\ref{subfig:mesonmass1}), (\ref{subfig:mesonmass2}) and (\ref{subfig:mesonmass3}) shown, at low density, i.e. $\mu<100$MeV and zero temperature, there is a 1st order gap at $\omega\simeq 0.8$GeV for the dependence of the chiral condensate on angular velocity.  After that the pion would break the constraint of Goldstone theorem, that is the mass increases to meet that of $\sigma$ meson which driven by the chiral symmetry. As the chemical potential increase further, the phase transition would be weaken into the crossover, and the mass dependence on the angular velocity would become more and more smooth as shown in Fig.(\ref{subfig:mesonmass2}) and (\ref{subfig:mesonmass3}).

\begin{figure}[!thb]
	\centering
	\includegraphics[width=0.65\textwidth]{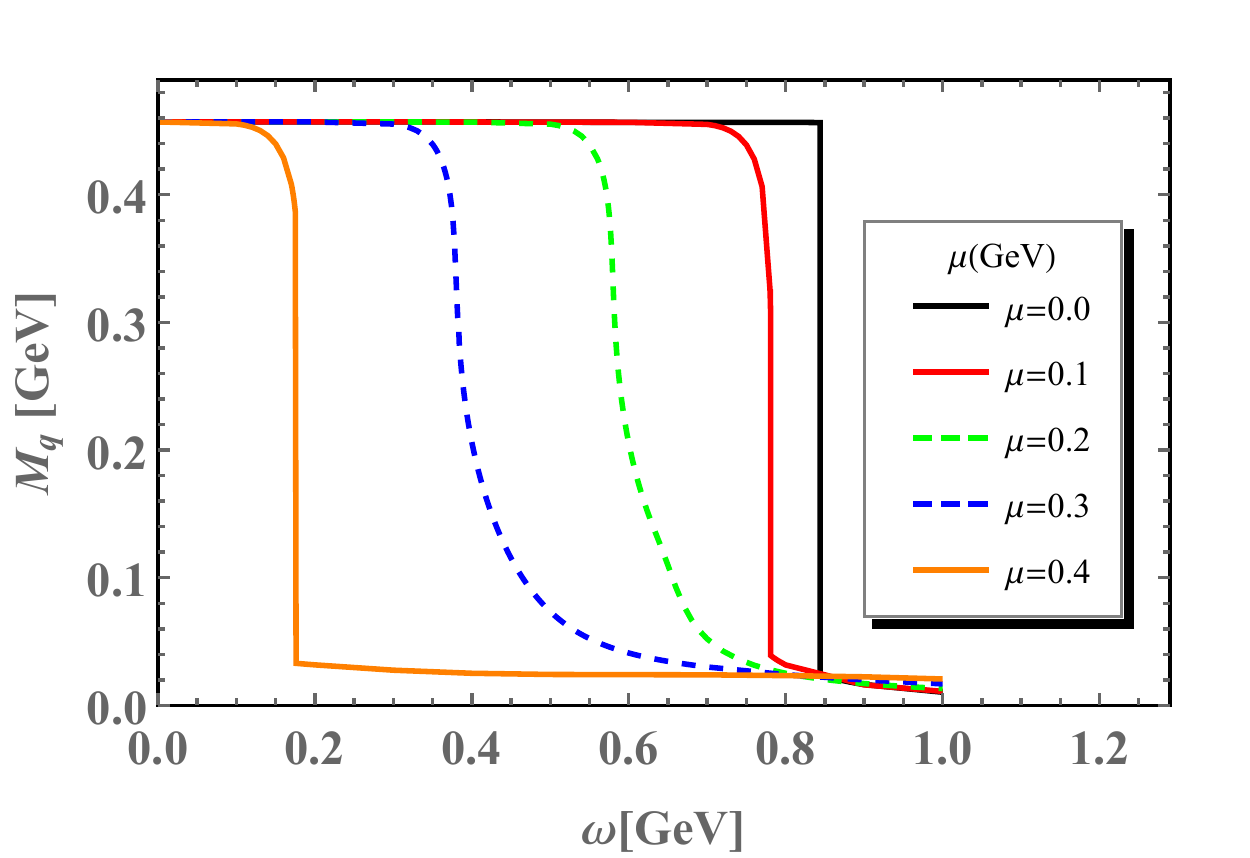}\\
	\caption{The constituent quark mass as a function of angular velocity for different chemical potentials. }
	\label{fig:quarkmass}	
\end{figure}

The constituent quark mass calculated from $M = m - 2G_S\left<\bar\psi\psi\right>$ as a function of angular velocity is shown in Fig.\ref{fig:quarkmass} for different chemical potentials. It is seen that the chiral condensate shows 1st order phase transition
at large angular velocity for small chemical potentials and at small angular velocity for large chemical potentials, this is in agreement
with the results in \cite{Wang:2018sur}, where it has been observed that the 1st order phase transition shows up in two corners of the 3D $T-\mu-\omega$ phase diagram.
	
From the numerical result it is clear that the angular velocity and chemical potential are complementary to each other when driven the chiral restoration. At low chemical potential the critical/crossover angular velocity is larger and become smaller when the chemical potential is larger.
However it is obvious that the chemical potential and angular velocity are not exactly equivalent to each other. Because physically the chemical potential is the energy shift from the difference between particle and anti-particle, while the shift induced by the rotation polarization is from the spin up and down difference. From this aspect the $\pm \frac{1}{2}\omega$ could be treated as the {\it spin chemical potential}. Analytically the difference could explicitly observed in the gap equation and the polarization functions as follows

\eq
\begin{aligned}
\Pi_{s}(0,i \nu_{n})=&N_{f}N_{c}\sum_{s=\pm}\int\frac{d^{3}\vec{p}}{(2\pi)^{3}}\left[
Res1(\vec{p},\nu_{n}) \theta \left(-\mu -\frac{s\omega }{2}+E_{p}\right) n_{f}\left(E_{p}-\mu -\frac{s\omega }{2},T\right)\right.\\
&+Res3(\vec{p},\nu_{n}) \theta \left(-\mu -\frac{s\omega }{2}+E_{p}\right) n_{f}\left(E_{p}-\mu -\frac{s\omega }{2},T\right)\\
&-Res1(\vec{p},\nu_{n}) \theta \left(\mu +\frac{s\omega }{2}-E_{p}\right) n_{f}\left(-E_{p}+\mu +\frac{s\omega }{2},T\right)
-Res2(\vec{p},\nu_{n}) n_{f}\left(E_{p}+\mu +\frac{s\omega }{2},T\right)\\
&-Res3(\vec{p},\nu_{n}) \theta \left(\mu +\frac{s\omega }{2}-E_{p}\right) n_{f}\left(-E_{p}+\mu +\frac{s\omega }{2},T\right)
-Res4(\vec{p},\nu_{n}) n_{f}\left(E_{p}+\mu +\frac{s\omega }{2},T\right)\\
&+Res1(\vec{p},\nu_{n}) \theta \left(\mu +\frac{s\omega }{2}-E_{p}\right)
+Res3(\vec{p},\nu_{n}) \theta \left(\mu +\frac{s\omega }{2}-E_{p}\right)\\
&-Res1\left(\vec{p},\nu_{n}\right)-Res3\left(\vec{p},\nu_{n}\right)
\Big],
\end{aligned}
\eeq

where $Res1(\vec{p},\nu_{n}),Res2(\vec{p},\nu_{n}),Res3(\vec{p},\nu_{n})$ and $Res4(\vec{p},\nu_{n})$ are residues in Eq.(\ref{residues1}).

And the polarization function $\Pi_{ps}$ is given as: 
\eq
\begin{aligned}
\Pi_{ps}(0,i \nu_{n})=&N_{f}N_{c}\sum_{s=\pm}\int\frac{d^{3}\vec{p}}{(2\pi)^{3}}\left[
Res1'(\vec{p},\nu_{n}) \theta \left(-\mu -\frac{s\omega }{2}+E_{p}\right) n_{f}\left(E_{p}-\mu -\frac{s\omega }{2},T\right)\right.\\
&+Res3'(\vec{p},\nu_{n}) \theta \left(-\mu -\frac{s\omega }{2}+E_{p}\right) n_{f}\left(E_{p}-\mu -\frac{s\omega }{2},T\right)\\
&-Res1'(\vec{p},\nu_{n}) \theta \left(\mu +\frac{s\omega }{2}-E_{p}\right) n_{f}\left(-E_{p}+\mu +\frac{s\omega }{2},T\right)
-Res2'(\vec{p},\nu_{n}) n_{f}\left(E_{p}+\mu +\frac{s\omega }{2},T\right)\\
&-Res3'(\vec{p},\nu_{n}) \theta \left(\mu +\frac{s\omega }{2}-E_{p}\right) n_{f}\left(-E_{p}+\mu +\frac{s\omega }{2},T\right)
-Res4'(\vec{p},\nu_{n}) n_{f}\left(E_{p}+\mu +\frac{s\omega }{2},T\right)\\
&+Res1'(\vec{p},\nu_{n}) \theta \left(\mu +\frac{s\omega }{2}-E_{p}\right)
+Res3'(\vec{p},\nu_{n}) \theta \left(\mu +\frac{s\omega }{2}-E_{p}\right)\\
&-Res1'(\vec{p},\nu_{n})-Res3'(\vec{p},\nu_{n})\Big].
\end{aligned}
\eeq
where $Res1'(\vec{p},\nu_{n}),Res2'(\vec{p},\nu_{n}),Res3'(\vec{p},\nu_{n})$ and $Res4'(\vec{p},\nu_{n})$ are residues in Eq.(\ref{residues2}).

It is clear that the functions depend on both the $\mu\pm\omega/2$ combinations. However it is also reasonable that the critical
behavior would take place at one of the angular velocitys which satisfy $E_p-\mu\pm\frac{\omega}{2}=0$. If we choose both the chemical
potential and angular velocity positive, the $\omega=2(m-\mu)$ part would dominate the critical behavior. Hence the chemical potential
and angular velocity appear to be complementary to each other on the determination of the critical point.

%the ref{}masses of meson suddenly changed with a first order phase transition. It is reasonable, because the rotation plays a role like effective chemical potential on quark mass. In our expression, the behavior of constituent quark mass made a difference with polarization function. After the integration for momentum $p$, there is no more other contribution from $\omega$.

%In different finite chemical potential case, the meson masses as a function of angular velocity $\omega$ also are shown in Fig.\ref{fig:mesonmass}. Although the rotation is interpreted as effective chemical potential by people~\cite{Chen:2015hfc}, two kind of "chemical potential" will not always promote each other.In Fig.\ref{subfig:mesonmass3} and Fig.\ref{subfig:mesonmass4},the pion mass started to ascend gently, while the $\sigma$ meson mass descends smoothly.

%When chemical potential is larger, the phase transition becomes first order again. In Fig.\ref{subfig:mesonmass5},$\mu=400MeV$, we can find the jump of meson mass. However, the behaviors of meson mass after chiral symmetry restoration are not normal. It descends firstly and ascends finally. Currently, we cannot interpret it with physical meaning.

\subsection{The $\rho$ meson}
\begin{figure}[t]
 	\includegraphics[width=0.65\textwidth]{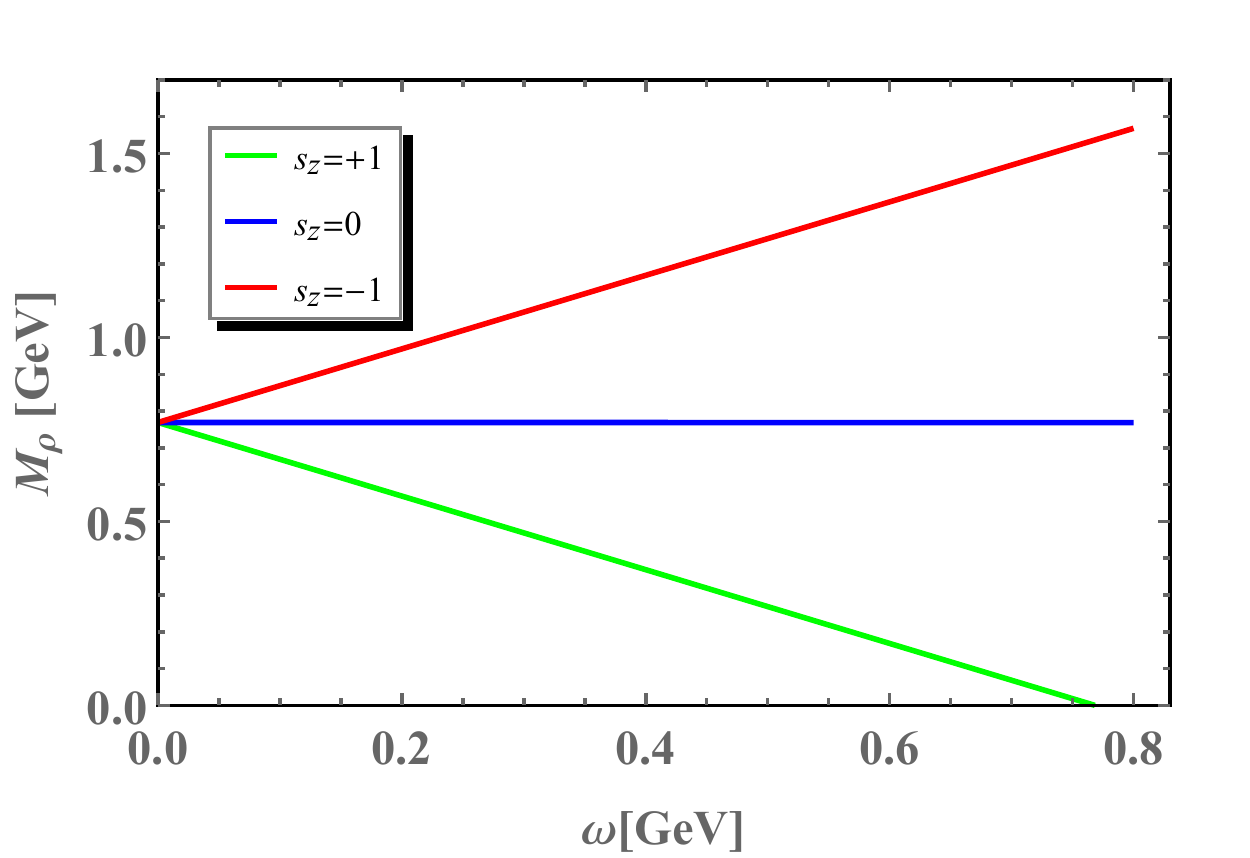}\\
 	\caption{$\rho$ meson masses as a function of angular velocity at temperature $T=10 MeV$.}\label{fig:rhomesonmass}
 \end{figure}

Taking the direction of rotation as the $z$-axies, and the three components of a massive vector meson can be represented as $s_{z}=\pm 1$ and $s_{z}=0$. And the nonzero spin ones would be polarized by the so-call Barnett effect which introduces the shift as $-{\vec\omega}\cdot{\vec S}$ to the energy levels under rotation. In our 2-flavor model we take the $\rho$ meson for example to explore the rotation-induced energy shift with the self-consistent numerical calculations at the quark level. Fig.\ref{fig:rhomesonmass} shows the numerical results for $\rho$ masses with $s_{z}=\pm 1$ and $s_{z}=0$ as functions of angular velocity at temperature $T=10$MeV. It is obvious that the splitting mass curves have shown the different influence of rotation. For the $s_z=0$ case there is no net angular momentum for the particle polarization by the rotation. This makes the mass dependence on the rotation is almost the same as the scalar case which stay invariant as the chiral condensate below the critical angular velocity.
While for the $s_z=\pm 1$ cases the rotation polarization would generate the energy shift $\mp \omega$ to the corresponding masses. This is confirmed by the numerical results in Fig.\ref{fig:rhomesonmass}. The mass dependence on the angular velocity is two straight lines for the $s_z=\pm 1$ components. The behavior could be analytically proven with explicit form of the polarization functions. In the pole approximation the masses are determined by the pole of the meson propagators as Eq.(\ref{vectorpole}). With straightforward computation in the Appendix the polarization functions of vector meson satisfy
\eq
\frac{1}{2}A_{1}^{2}(m_{\rho}+\omega)+\frac{1}{2}A_{2}^{2}(m_{\rho}-\omega)
=A_{3}^{2}(m_{\rho}).
\eeq
This means the $m_\rho(\omega=0)-s_z\omega$ are exactly the masses of $s_z=0,\pm 1$ components. The mass of $s_z=1$ spin component decreases linearly with the angular velocity, and reaches zero at the critical angular velocity $\omega_c=m_\rho(\omega=0)$. Beyond the critical angular velocity $\omega_c$, the $s_z=1$ spin component of vector meson will develop condensation in the vacuum and this indicates that the system will be spontaneously spin polarized under strong rotation.

%For $S_{z}=+1$, the mass of $\rho$ meson decreases smoothly with the angular velocity, although we know that the quark mass is increasing with the rotation. At $\omega=0.77 GeV$, the $\rho$ meson mass drops to zero, and it may imply a $\rho$ condensation. On the contrary, for $S_{z}=-1$, the mass of  $\rho$ meson increases slowly but observably with the angular velocity growing. For $S_{z}=0$, the mass of $\rho$ meson stay at a platform, which means that the rotation effect is of little account.

\section{Conclusion}
\label{Conclusion}
Using the NJL model with vector channel interaction we have calculated the scalar, pseudoscalar and vector mesons' masses at finite temperature, chemical potential and angular velocity. In the RPA and pole approximation the mesons are treated as the effective degree of freedoms which transmit the interaction between quarks. And the masses are determined by the polarization functions. This approximation could preserve the Goldstone theorem explicitly although
the back reaction of meson to the phase transition is neglected. Because of the four-fermion point interaction and pole approximation the microscopic details of mesons have been lost. And all of them behave as fundamental particles which are polarized by rotation according to their net spin angular momentum. For the scalar and pseudoscalar cases the mass spectra are controlled by the chiral condensate which is the main mechanism generating the hadron mass in NJL model.
At low temperature and chemical potential the chiral restoration is 1st order  which make the meson masses a sudden jump at the critical angular velocity. While as the temperature or chemical potential increasing the phase transition would degenerate to crossovers which also smoothen the mass curves of mesons along the angular velocity.  It is easy to expect that at large enough angular velocity the vector condensate vacuum would be preferred and the corresponding effective mass should be zero. That is why we have only studied the vector meson's mass behavior below the $\omega=m_\rho$. It is found that although the polarization function computation is complicated masses of the three components $s_z=0, \pm 1$ are the same as the result by treating them as the fundamental particles, that is $m_\rho$ and $m_\rho\pm \omega$.  Once the chiral restored the vector condensate would emerge simultaneously which will be studied in our next work.

In non-central heavy-ion collisions, the created system carries large angular momentum. The properties of particles will be changed under rotating medium. In this paper, we investigated the behavior of scalar and vector meson mass under the rotation.
It is found that the behavior of scalar and pseudoscalar meson masses under the angular velocity $\omega$ is similar to that at finite chemical potential, both rely on the behavior of constituent quark mass and reflect the property related to the chiral symmetry. However, masses of vector meson have more profound relation with rotation. After tedious calculation, it turns out that at low temperature and small chemical potenial, the mass for spin component $s_z=0,\pm 1$ of vector meson under rotation shows very simple mass splitting relation $m_{\rho}^{s_z}(\omega)=m_\rho(\omega=0)-\omega s_z$, similar to the Zeeman splitting of charged meson under magnetic fields. Especially it is noticed that the mass of spin component $s_z=1$ vector meson $\rho$ decreases linearly with $\omega$ and reaches zero at $\omega_c=m_\rho(\omega=0)$, this indicates the system will develop $s_z=1$ vector meson condensation and the system will be spontaneously spin polarized under rotation. It deserves further study to compare the spin polarization with $s_z=1$ vector meson condensation and the spin polarization defined by the
condensation of $<{\bar\psi}i\sigma^{\mu\nu}\psi>$ proposed in \cite{Tatsumi-spinpolarization}.

%For scalar meson, the behavior of meson mass is contributed from the behavior of constituent quark mass. For vector meson, the rotation really makes a difference with mass structure of vector meson. Three splitting branches are shown in $M_{\rho}-\omega$ relation. In addition, the mass of $\rho$ meson with spin component $s_{z}=+1$ will drop to zero. it maybe imply a $\rho$ condensation.

\begin{acknowledgements}
	We thank Kun Xu for useful discussion. M.H.is supported by the NSFC under Grant Nos. 11725523 and 11735007,  Chinese Academy of Sciences under Grant No. XDPB09, the start-up funding from University of Chinese Academy of Sciences(UCAS), and the Fundamental Research Funds for the Central Universities .	Y.J. is  supported by NSFC under Grant No.11875002 and the Zhuobai Program of Beihang University.
\end{acknowledgements}

\appendix
 \section{The polarization function under rotation}
 \label{appendix:a}
 For the $\sigma$ meson, substituting the Eq.~\ref{propagator} into the definition of polarization function, the scalar proper polarization function under rotation is given as
 \eq
\begin{aligned}
\Pi_{s}(q)&=-i \int d^{4}\tilde{r}Tr_{sfc}\left[i S(0;\tilde{r})i S(\tilde{r};0)\right]e^{i q\cdot \tilde{r}}\\
&=-i N_{f} N_{c} \sum_{n=-\infty}^{+\infty}\sum_{l=-\infty}^{+\infty}\int d^{4}\tilde{r}
\int \frac{d k_{0}d k_{z}}{(2\pi)^{2}}\int_{0}^{+\infty}\frac{k_{t}d k_{t}}{2\pi}
\int \frac{d p_{0}d p_{z}}{(2\pi)^{2}} \int_{0}^{+\infty}\frac{p_{t}d  p_{t}}{2\pi} \\
&\times Tr(A_{n}B_{l})\times \frac{e^{-ik_{0}t+ik_{z}z}e^{n\phi}}{[k_{0}+(n+\frac{1}{2})\omega]^{2}-k_{t}^{2}-k_{z}^{2}-M^{2}+i\epsilon}
\times \frac{e^{ip_{0}t-ip_{z}z}e^{-l\phi}}{[p_{0}+(l+\frac{1}{2})\omega]^{2}-p_{t}^{2}-p_{z}^{2}-M^{2}+i\epsilon}\times e^{i q\cdot \tilde{r}},
\end{aligned}
\eeq
where
\eq
\begin{aligned}
A_{n}&=
\begin{tiny}
 \begin{pmatrix}
        \begin{smallmatrix}
 \left(k_{0}+M+\left(n+\frac{1}{2}\right) \omega \right) J_n(k_t r) J_n(0) & 0 & -k_z J_n(k_t r) J_n(0) & i k_t J_n(k_t r) J_{n+1}(0) \\
 0 & \left(k_0+M+\left(n+\frac{1}{2}\right) \omega \right) e^{i \phi } J_{n+1}(k_{t} r) J_{n+1}(0) & -i e^{i \phi } k_{t} J_{n+1}(k_{t} r) J_n(0) & k_{z} e^{i \phi } J_{n+1}(k_{t} r) J_{n+1}(0) \\
 k_{z} J_n(k_{t} r) J_n(0) & -i k_{t} J_n(k_{t} r) J_{n+1}(0) & -\left(k_{0}-M+\left(n+\frac{1}{2}\right) \omega \right) J_n(k_{t} r) J_n(0) & 0 \\
 i e^{i \phi } k_{t} J_{n+1}(k_{t} r) J_n(0) & -k_{z} e^{i \phi } J_{n+1}(k_{t} r) J_{n+1}(0) & 0 & -\left(k_{0}-M+\left(n+\frac{1}{2}\right) \omega \right) e^{i \phi } J_{n+1}(k_{t} r) J_{n+1}(0)\\
\end{smallmatrix}
\end{pmatrix},
\end{tiny}\\
B_{l}&=
\begin{tiny}
 \begin{pmatrix}
 \begin{smallmatrix}
 \left(M+p_{0}+\left(l+\frac{1}{2}\right) \omega \right) J_l(p_{t} r) J_l(0) & 0 & -p_{z} J_l(p_{t} r) J_l(0) & i p_{t} e^{-i \phi } J_{l+1}(p_{t} r) J_l(0) \\
 0 & \left(M+p_{0}+\left(l+\frac{1}{2}\right) \omega \right) e^{-i \phi } J_{l+1}(p_{t} r) J_{l+1}(0) & -i p_{t} J_l(p_{t} r) J_{l+1}(0) & p_{z} e^{-i \phi } J_{l+1}(p_{t} r) J_{l+1}(0)\\
 p_{z} J_l(p_{t} r) J_l(0) & -i p_{t} e^{-i \phi } J_{l+1}(p_{t} r) J_l(0) & -\left(-M+p_{0}+\left(l+\frac{1}{2}\right) \omega \right) J_l(p_{t} r) J_l(0) & 0\\
 i p_{t} J_l(p_{t} r) J_{l+1}(0) & -p_{z} e^{-i \phi } J_{l+1}(p_{t} r) J_{l+1}(0) & 0 & -\left(-M+p_{0}+\left(l+\frac{1}{2}\right) \omega \right) e^{-i \phi } J_{l+1}(p_{t} r) J_{l+1}(0)\\
\end{smallmatrix}
\end{pmatrix}.
\end{tiny}
\end{aligned}
\eeq

Here we give a matrix form instead of the summation of projection operators in Eq.~\ref{propagator}. The symbol "$Tr_{sfc}$" stands for evaluate the trace on spinor, flavor and color space. After a tedious calculation, we get
\eq
\begin{aligned}
Tr(A_{n}B_{l})&=[\frac{1}{2} (2 k_{0}+2 n \omega +\omega ) (2 l \omega +2 p_{0}+\omega )+2 M^2
-2 k_{z} p_{z}] J_l(0) J_n(0) J_n(k_{t} r) J_l(p_{t} r)\\
&+[\frac{1}{2} (2 k_{0}+2 n \omega +\omega ) (2 l \omega +2 p_{0}+\omega )+2 M^2
-2 k_{z} p_{z}]J_{l+1}(0) J_{n+1}(0) J_{n+1}(k_{t} r) J_{l+1}(p_{t} r)\\
&-2 k_{t} p_{t} J_{l+1}(0) J_{n+1}(0) J_n(k_{t} r) J_l(p_{t} r)\\
&-2 k_{t} p_{t} J_l(0) J_n(0) J_{n+1}(k_{t} r) J_{l+1}(p_{t} r).
\end{aligned}
\eeq

When $n\neq 0$, it's obvious that $J_{n}(0)=0$ and $J_{0}(0)=1$. As a consequence, the result of the summation will have finite terms
\eq
\begin{aligned}
\Pi_{s}(q)&=-i \int d^{4}\tilde{r}Tr_{sfc}\left[S(0;\tilde{r})S(\tilde{r};0)\right]e^{i q\cdot \tilde{r}}\\
&=-i N_{f}N_{c}\int d^{4}\tilde{r}
\int \frac{d k_{0}d k_{z}}{(2\pi)^{2}}\int_{0}^{+\infty}\frac{k_{t}d k_{t}}{2\pi}
\int \frac{d p_{0}d p_{z}}{(2\pi)^{2}} \int_{0}^{+\infty}\frac{p_{t}d  p_{t}}{2\pi} \\
&\times
\left\{ \left[\frac{1}{2} (2 k_{0}+\omega ) ( 2 p_{0}+\omega )+2 M^{2}-2 k_{z} p_{z}\right]
\times
\frac{J_{0}(k_{t} r)J_{0}(p_{t}r)e^{-i k_{0}t+i k_{z}z}e^{i p_{0}t-i p_{z}z}}{\left[(k_{0}+\frac{1}{2}\omega)^{2}-k_{t}^{2}-k_{z}^{2}-M^{2}+i\epsilon\right]
\left[(p_{0}+\frac{1}{2}\omega)^{2}-p_{t}^{2}-p_{z}^{2}-M^{2}+i\epsilon\right]}\right.\\
&+\left[\frac{1}{2} (2 k_{0}-\omega ) ( 2 p_{0}-\omega )+2 M^{2}-2 k_{z} p_{z}\right]\times\frac{J_{0}(k_{t} r)J_{0}(p_{t}r)e^{-i k_{0}t+i k_{z}z}e^{i p_{0}t-i p_{z}z}}{\left[(k_{0}-\frac{1}{2}\omega)^{2}-k_{t}^{2}-k_{z}^{2}-M^{2}+i\epsilon\right]
\left[(p_{0}-\frac{1}{2}\omega)^{2}-p_{t}^{2}-p_{z}^{2}-M^{2}+i\epsilon\right]}\\
&-2k_{t}p_{t}\times\frac{J_{1}(k_{t} r)J_{1}(p_{t}r)e^{-i k_{0}t+i k_{z}z}e^{i p_{0}t-i p_{z}z}}{\left[(k_{0}+\frac{1}{2}\omega)^{2}-k_{t}^{2}-k_{z}^{2}-M^{2}+i\epsilon\right]
\left[(p_{0}+\frac{1}{2}\omega)^{2}-p_{t}^{2}-p_{z}^{2}-M^{2}+i\epsilon\right]}\\
&-2k_{t}p_{t}\times\left.\frac{J_{-1}(k_{t} r)J_{-1}(p_{t}r)e^{-i k_{0}t+i k_{z}z}e^{i p_{0}t-i p_{z}z}}{\left[(k_{0}-\frac{1}{2}\omega)^{2}-k_{t}^{2}-k_{z}^{2}-M^{2}+i\epsilon\right]
\left[(p_{0}-\frac{1}{2}\omega)^{2}-p_{t}^{2}-p_{z}^{2}-M^{2}+i\epsilon\right]}
\right\}\times e^{i q\cdot \tilde{r}}.
\end{aligned}
\eeq
Applying the integral representation of Bessel functions, the polarization function can be simplified. In integral representation, Bessel functions are expressed as:
\eq
\begin{aligned}
J_{n}(r)&=\frac{1}{2\pi}\int_{-\pi}^{+\pi}e^{i(r\sin\theta-n\theta)}d\theta,\\
J_{0}(r)&=\frac{1}{2\pi}\int_{0}^{2\pi}e^{\pm i r\cos\theta}d\theta,\\
J_{1}(r)&=\frac{1}{2\pi i}\int_{0}^{2\pi}e^{i r\cos\theta\pm i\theta}d\theta,\\
J_{1}(r)&=-\frac{1}{2\pi i}\int_{0}^{2\pi}e^{-i r\cos\theta\pm i\theta}d\theta.
\end{aligned}
\eeq
Let $\vec{k_{t}}=(k_{t},\phi+\theta)=(k_{x},k_{y}),\vec{r}=(r,\phi)=(x,y)$, and then we have the transformation formulae:
\eq
\int_{0}^{\infty}\frac{k_{t}d k_{t}}{2\pi}\int_{0}^{2\pi}\frac{d\theta}{2\pi i}i k_{t}e^{i \phi}e^{i k_{t}r \cos \theta+i \theta}
=\int \frac{d k_{x}d k_{y}}{(2\pi)^2}(k_{x}+i k_{y})e^{i\vec{k_{t}}\cdot \vec{r}},
\eeq
\eq
\int_{0}^{\infty}\frac{k_{t}d k_{t}}{2\pi}\int_{0}^{2\pi}\frac{d\theta}{2\pi i}i k_{t}e^{-i \phi}e^{-i k_{t}r \cos \theta-i \theta}
=\int \frac{d k_{x}d k_{y}}{(2\pi)^2}(k_{x}-i k_{y})e^{-i\vec{k_{t}}\cdot \vec{r}}.
\eeq
Applied the transformation formulae, the polarization function for scalar meson can be expressed without Bessel function. It will be more efficient for numerical calculation.
\eq
\begin{aligned}
\Pi_{s}(q)&=-i N_{f}N_{c}\int d^{4}\tilde{r}
\int \frac{d^{4} k}{(2\pi)^{4}}
\int \frac{d^{4} p}{(2\pi)^{4}} \\
&\times
\left\{ \left[\frac{1}{2} (2 k_{0}+\omega ) ( 2 p_{0}+\omega )+2 M^{2}-2 k_{z} p_{z}\right]
\times
\frac{e^{-i k\cdot\tilde{r}}e^{i p\cdot\tilde{r}}}{\left[\left(k_{0}+\frac{1}{2}\omega\right)^{2}-\vec{k}^{2}-M^{2}+i\epsilon\right]
\left[\left(p_{0}+\frac{1}{2}\omega\right)^{2}-\vec{p}^{2}-M^{2}+i\epsilon\right]}\right.\\
&+ \left[\frac{1}{2} (2 k_{0}-\omega )( 2 p_{0}-\omega )+2 M^{2}-2 k_{z} p_{z}\right]\times\frac{e^{-i k\cdot\tilde{r}}e^{i p\cdot\tilde{r}}}{\left[(k_{0}-\frac{1}{2}\omega)^{2}-\vec{k}^{2}-M^{2}+i\epsilon\right]
\left[\left(p_{0}-\frac{1}{2}\omega\right)^{2}-\vec{p}^{2}-M^{2}+i\epsilon\right]}\\
&-2(k_{x}+i k_{y})(p_{x}-i p_{y})\times
\frac{e^{-i k\cdot\tilde{r}}e^{i p\cdot\tilde{r}}}{\left[\left(k_{0}+\frac{1}{2}\omega\right)^{2}-\vec{k}^{2}-M^{2}+i\epsilon\right]
\left[\left(p_{0}+\frac{1}{2}\omega\right)^{2}-\vec{p}^{2}-M^{2}+i\epsilon\right]}\\
&-2(k_{x}+i k_{y})(p_{x}-i p_{y})\times
\frac{e^{-i k\cdot\tilde{r}}e^{i p\cdot\tilde{r}}}{\left[\left(k_{0}-\frac{1}{2}\omega\right)^{2}-\vec{k}^{2}-M^{2}+i\epsilon\right]
\left[\left(p_{0}-\frac{1}{2}\omega\right)^{2}-\vec{p}^{2}-M^{2}+i\epsilon\right]}
\}\times e^{i q\cdot \tilde{r}}.\\
\end{aligned}
\eeq
Furthermore, integrating the $\tilde{r}$ and $k$ analytically, we can get the polarization function as following:
\eq
\begin{aligned}
\Pi_{s}(q)&=
-i N_{f}N_{c}\int \frac{d^{4} p}{(2\pi)^{4}} \\
&\times
\left\{
\frac{\left[2 \left(p_{0}+q_{0}+\frac{1}{2}\omega \right) \left(  p_{0}+\frac{1}{2}\omega \right)+2 M^{2}-2 (p_{z}+q_{z}) p_{z}\right]-2\left[(p_{x}+q_{x})+i (p_{y}+q_{y})\right](p_{x}-i p_{y})}
{\left[\left(p_{0}+q_{0}+\frac{1}{2}\omega\right)^{2}-(\vec{p}+\vec{q})^{2}-M^{2}+i\epsilon\right]
\left[\left(p_{0}+\frac{1}{2}\omega\right)^{2}-\vec{p}^{2}-M^{2}+i\epsilon\right]}\right.\\
&+ \left.\frac{\left[2 \left(p_{0}+q_{0}-\frac{1}{2}\omega \right) \left( p_{0}-\frac{1}{2}\omega \right)+2 M^{2}-2 (p_{z}+q_{z}) p_{z}\right]-2\left[(p_{x}+q_{x})+i (p_{y}+q_{y})\right](p_{x}-i p_{y})}
{\left[\left(p_{0}+q_{0}-\frac{1}{2}\omega\right)^{2}-(\vec{p}+\vec{q})^{2}-M^{2}+i\epsilon\right]
\left[\left(p_{0}-\frac{1}{2}\omega\right)^{2}-\vec{p}^{2}-M^{2}+i\epsilon\right]}
\right\}.
\end{aligned}
\eeq
Due to symmetric analysis for integration, the expression can be simplified as following:
\eq
\begin{aligned}
\Pi_{s}(q^2)&=
-2 i N_{f}N_{c}\int \frac{d^{4} p}{(2\pi)^{4}} \\
&\times
\left\{ \frac{\left(p_{0}+q_{0}+\frac{1}{2}\omega \right) \left(p_{0}+\frac{1}{2}\omega \right)+M^2-(\vec{p}+\vec{q}) \vec{p}}
{\left[\left(p_{0}+q_{0}+\frac{1}{2}\omega\right)^{2}-(\vec{p}+\vec{q})^{2}-M^{2}\right]
\left[\left(p_{0}+\frac{1}{2}\omega\right)^{2}-\vec{p}^{2}-M^{2}\right]}\right.\\
&+ \left.\frac{\left(p_{0}+q_{0}-\frac{1}{2}\omega \right) \left(p_{0}-\frac{1}{2}\omega \right)+M^2-(\vec{p}+\vec{q}) \vec{p}}
{\left[\left(p_{0}+q_{0}-\frac{1}{2}\omega\right)^{2}-(\vec{p}+\vec{q})^{2}-M^{2}\right]
\left[\left(p_{0}-\frac{1}{2}\omega\right)^{2}-\vec{p}^{2}-M^{2}\right]}
\right\}.
\end{aligned}
\eeq
For finite temperature formalism:
\eq
p_{0}\rightarrow i \tilde{\omega}_{N},\hspace{10pt}
q_{0}\rightarrow i \nu_{n},\hspace{10pt}
\int \frac{p_{0}}{2\pi}\rightarrow i T\sum_{N},\hspace{10pt}
\tilde{\omega}_{N}=(2N+1)\pi T.
\eeq
The polarization function at finite temperature and chemical potential under rotation can be rewritten as:
\begin{equation}
	\Pi_{s}(\vec{q},i\nu_{n})=2N_{f} N_{c}T \sum_{s=\pm}\sum_N\int \frac{d^{3}\vec{p}}{(2\pi)^{3}}
	\frac{\left[(i \tilde{\omega}_{N}+i \nu_{n})+\frac{1}{2}s\omega+\mu\right]\left(i \tilde{\omega}_{N}+\frac{1}{2}s\omega+\mu\right)+M^{2}-(\vec{p}+\vec{q})\cdot\vec{p}}{\left[\left(i \tilde{\omega}_{N}+i \nu_{n}+\frac{1}{2}s\omega+\mu\right)^{2}-(\vec{p}+\vec{q})^{2}-M^{2}\right]\left[\left(i \tilde{\omega}_{N}+\frac{1}{2}s\omega+\mu\right)^{2}-\vec{p}^{2}-M^{2}\right]}.
\end{equation}
Setting $\vec{q}=0$, Matsubara Summation will give us a result in term of residue theorem:
\eq
\begin{aligned}
\Pi_{s}(0,i \nu_{n})=&N_{f}N_{c}\sum_{s=\pm}\int\frac{d^{3}\vec{p}}{(2\pi)^{3}}\left[
Res1(\vec{p},\nu_{n}) \theta \left(-\mu -\frac{s\omega }{2}+E_{p}\right) n_{f}\left(E_{p}-\mu -\frac{s\omega }{2},T\right)\right.\\
&+Res3(\vec{p},\nu_{n}) \theta \left(-\mu -\frac{s\omega }{2}+E_{p}\right) n_{f}\left(E_{p}-\mu -\frac{s\omega }{2},T\right)\\
&-Res1(\vec{p},\nu_{n}) \theta \left(\mu +\frac{s\omega }{2}-E_{p}\right) n_{f}\left(-E_{p}+\mu +\frac{s\omega }{2},T\right)
-Res2(\vec{p},\nu_{n}) n_{f}\left(E_{p}+\mu +\frac{s\omega }{2},T\right)\\
&-Res3(\vec{p},\nu_{n}) \theta \left(\mu +\frac{s\omega }{2}-E_{p}\right) n_{f}\left(-E_{p}+\mu +\frac{s\omega }{2},T\right)
-Res4(\vec{p},\nu_{n}) n_{f}\left(E_{p}+\mu +\frac{s\omega }{2},T\right)\\
&+Res1(\vec{p},\nu_{n}) \theta \left(\mu +\frac{s\omega }{2}-E_{p}\right)
+Res3(\vec{p},\nu_{n}) \theta \left(\mu +\frac{s\omega }{2}-E_{p}\right)\\
&-Res1\left(\vec{p},\nu_{n}\right)-Res3\left(\vec{p},\nu_{n}\right)
\Big],
\end{aligned}
\eeq
where $n_{f}(x,T)=\frac{1}{e^{x/T}+1}$ is distribution function, and four residues are given as following:
\eq
\label{residues1}
\begin{aligned}
    Res1(\vec{p},\nu_{n})=&\frac{-i \nu_{n} E_{p}+E_{p}^{2}+M^{2}-\vec{p}^{2}}{E_{p} \left(-\nu_{n}^2-2 i \nu_{n} E_{p}\right)},\\
    Res2(\vec{p},\nu_{n})=&\frac{-i \nu_{n} E_{p}-E_{p}^{2}-M^{2}+\vec{p}^{2}}{E_{p} \left(-\nu_{n}^2+2 i \nu_{n} E_{p}\right)},\\
    Res3(\vec{p},\nu_{n})=&\frac{+i \nu_{n} E_{p}+E_{p}^{2}+M^{2}-\vec{p}^{2}}{E_{p} \left(-\nu_{n}^2+2 i \nu_{n} E_{p}\right)},\\
    Res4(\vec{p},\nu_{n})=&\frac{i \nu_{n} E_{p}-E_{p}^{2}-M^{2}+\vec{p}^{2}}{E_{p} \left(-\nu_{n}^2-2 i \nu_{n} E_{p}\right)}.
\end{aligned}
\eeq
We should notice that $E_{\bm{p}}=\sqrt{\bm{p}^{2}+M^{2}}$ and quark mass $M$ is a function of angular velocity $\omega$. For psuadoscalar meson, the finite temperature version polarization function is:
\begin{equation}
	\Pi_{ps}(\vec{q},i\nu_{n})=-4N_{f} N_{c}T \sum_{s=\pm}\sum_N\int \frac{d^{3}\vec{p}}{(2\pi)^{3}}
	\frac{[(i \tilde{\omega}_{N}+i \nu_{n})+\frac{1}{2}s\omega+\mu][i \tilde{\omega}_{N}+\frac{1}{2}s\omega+\mu]-M^{2}-(\vec{p}+\vec{q})\cdot\vec{p}}{[(i \tilde{\omega}_{N}+i \nu_{n}+\frac{1}{2}s\omega+\mu)^{2}-(\vec{p}+\vec{q})^{2}-M^{2}][(i \tilde{\omega}_{N}+\frac{1}{2}s\omega+\mu)^{2}-\vec{p}^{2}-M^{2}]}.
\end{equation}
Seting $\vec{q}=0$, Matsubara Summation gives:
\eq
\begin{aligned}
\Pi_{ps}(0,i \nu_{n})=&N_{f}N_{c}\sum_{s=\pm}\int\frac{d^{3}\vec{p}}{(2\pi)^{3}}\left[
Res1'(\vec{p},\nu_{n}) \theta \left(-\mu -\frac{s\omega }{2}+E_{p}\right) n_{f}\left(E_{p}-\mu -\frac{s\omega }{2},T\right)\right.\\
&+Res3'(\vec{p},\nu_{n}) \theta \left(-\mu -\frac{s\omega }{2}+E_{p}\right) n_{f}\left(E_{p}-\mu -\frac{s\omega }{2},T\right)\\
&-Res1'(\vec{p},\nu_{n}) \theta \left(\mu +\frac{s\omega }{2}-E_{p}\right) n_{f}\left(-E_{p}+\mu +\frac{s\omega }{2},T\right)
-Res2'(\vec{p},\nu_{n}) n_{f}\left(E_{p}+\mu +\frac{s\omega }{2},T\right)\\
&-Res3'(\vec{p},\nu_{n}) \theta \left(\mu +\frac{s\omega }{2}-E_{p}\right) n_{f}\left(-E_{p}+\mu +\frac{s\omega }{2},T\right)
-Res4'(\vec{p},\nu_{n}) n_{f}\left(E_{p}+\mu +\frac{s\omega }{2},T\right)\\
&+Res1'(\vec{p},\nu_{n}) \theta \left(\mu +\frac{s\omega }{2}-E_{p}\right)
+Res3'(\vec{p},\nu_{n}) \theta \left(\mu +\frac{s\omega }{2}-E_{p}\right)\\
&-Res1'(\vec{p},\nu_{n})-Res3'(\vec{p},\nu_{n})\Big],
\end{aligned}
\eeq
where
\eq
\label{residues2}
\begin{aligned}
    Res1'(\vec{p},\nu_{n})=&\frac{i \nu_{n} E_{p}+E_{p}^{2}+M^{2}+\vec{p}^{2}}{E_{p} \left(-\nu_{n}^2-2 i \nu_{n} E_{p}\right)},\\
    Res2'(\vec{p},\nu_{n})=&\frac{i \nu_{n} E_{p}+E_{p}^{2}-M^{2}-\vec{p}^{2}}{E_{p} \left(-\nu_{n}^2+2 i \nu_{n} E_{p}\right)},\\
    Res3'(\vec{p},\nu_{n})=&\frac{-i \nu_{n} E_{p}-E_{p}^{2}+M^{2}+\vec{p}^{2}}{E_{p} \left(-\nu_{n}^2+2 i \nu_{n} E_{p}\right)},\\
    Res4'(\vec{p},\nu_{n})=&\frac{-i \nu_{n} E_{p}+E_{p}^{2}-M^{2}-\vec{p}^{2}}{E_{p} \left(-\nu_{n}^2-2 i \nu_{n} E_{p}\right)}.
\end{aligned}
\eeq

\section{The polarization function for vector meson under rotation}
\label{appendix:b}
For $\rho$ meson, the polarization function with one loop contribution can be expressed as
\eq
\Pi^{\mu\nu,ab}=-i \int d^{4}\tilde{r}Tr_{sfc}\left[i \gamma^{\mu}\tau^{a}S(0;\tilde{r})i \gamma^{\nu}\tau^{b}S(\tilde{r};0)\right]e^{i q\cdot \tilde{r}}.
\eeq
Using the approach introduced in Appendix A. It is obvious that the charge of $\rho$ meson will make on difference with polarization function under rotation. We can get the nonzero elements of the matrix
\eq
\Pi^{\mu\nu}_{\rho}=\left(
                      \begin{array}{cccc}
                        0 & 0 & 0 & 0 \\
                        0 & \Pi^{11} & \Pi^{12} & 0 \\
                        0 & \Pi^{21} & \Pi^{22} & 0 \\
                        0 & 0 & 0 & \Pi^{33} \\
                      \end{array}
                    \right).
\eeq
Using the same method in Appendix~\ref{appendix:a} and setting $\vec{q}=0$, we will get the nonzero elements which is given by:
\eq
\begin{aligned}
\Pi^{11}(q_{0})&=
N_{f}N_{c}\int \frac{d^{4} p}{(2\pi)^{4}} \\
&\times
\left\{-\frac{2 M^2-2 \left(p_{0}+\frac{\omega}{2}\right) \left(p_{0}+q_{0}-\frac{\omega}{2}\right)-2 p_{x}^2+2 p_{y}^2+2 p_{z}^2}
{[\left(p_{0}+\frac{\omega }{2}\right)^2-\vec{p}^2-M^2] [\left(p_{0}+q_{0}-\frac{\omega }{2}\right)^2-\vec{p}^2-M^2]}
-\frac{2 M^2-2 \left(p_{0}-\frac{\omega }{2}\right) \left(p_{0}+q_{0}+\frac{\omega }{2}\right)-2 p_{x}^2+2 p_{y}^2+2 p_{z}^2}
{[\left(p_{0}-\frac{\omega }{2}\right)^2-\vec{p}^2-M^2] [\left(p_{0}+q_{0}+\frac{\omega }{2}\right)^2-\vec{p}^2-M^2]}\right\},
\end{aligned}
\eeq

\eq
\begin{aligned}
\Pi^{12}(q_{0})&=
-i N_{f}N_{c}\int \frac{d^{4} p}{(2\pi)^{4}} \\
&\times
\left\{\frac{2 M^2-2 \left(p_{0}+\frac{\omega}{2}\right) \left(p_{0}+q_{0}-\frac{\omega}{2}\right)-2 p_{x}^2+2 p_{y}^2+2 p_{z}^2}
{[\left(p_{0}+\frac{\omega }{2}\right)^2-\vec{p}^2-M^2] [\left(p_{0}+q_{0}-\frac{\omega }{2}\right)^2-\vec{p}^2-M^2]}
-\frac{2 M^2-2 \left(p_{0}-\frac{\omega }{2}\right) \left(p_{0}+q_{0}+\frac{\omega }{2}\right)-2 p_{x}^2+2 p_{y}^2+2 p_{z}^2}
{[\left(p_{0}-\frac{\omega }{2}\right)^2-\vec{p}^2-M^2] [\left(p_{0}+q_{0}+\frac{\omega }{2}\right)^2-\vec{p}^2-M^2]}\right\},
\end{aligned}
\eeq

\eq
\begin{aligned}
\Pi^{21}(q_{0})&=
i N_{f}N_{c}\int \frac{d^{4} p}{(2\pi)^{4}} \\
&\times
\left\{\frac{2 M^2-2 \left(p_{0}+\frac{\omega}{2}\right) \left(p_{0}+q_{0}-\frac{\omega}{2}\right)+2 p_{x}^2-2 p_{y}^2+2 p_{z}^2}
{[\left(p_{0}+\frac{\omega }{2}\right)^2-\vec{p}^2-M^2] [\left(p_{0}+q_{0}-\frac{\omega }{2}\right)^2-\vec{p}^2-M^2]}
-\frac{2 M^2-2 \left(p_{0}-\frac{\omega }{2}\right) \left(p_{0}+q_{0}+\frac{\omega }{2}\right)+2 p_{x}^2-2 p_{y}^2+2 p_{z}^2}
{[\left(p_{0}-\frac{\omega }{2}\right)^2-\vec{p}^2-M^2] [\left(p_{0}+q_{0}+\frac{\omega }{2}\right)^2-\vec{p}^2-M^2]}\right\},
\end{aligned}
\eeq

\eq
\begin{aligned}
\Pi^{22}(q_{0})&=
-N_{f}N_{c}\int \frac{d^{4} p}{(2\pi)^{4}} \\
&\times
\left\{\frac{2 M^2-2 \left(p_{0}+\frac{\omega}{2}\right) \left(p_{0}+q_{0}-\frac{\omega}{2}\right)-2 p_{x}^2+2 p_{y}^2+2 p_{z}^2}
{[\left(p_{0}+\frac{\omega }{2}\right)^2-\vec{p}^2-M^2] [\left(p_{0}+q_{0}-\frac{\omega }{2}\right)^2-\vec{p}^2-M^2]}
+\frac{2 M^2-2 \left(p_{0}-\frac{\omega }{2}\right) \left(p_{0}+q_{0}+\frac{\omega }{2}\right)-2 p_{x}^2+2 p_{y}^2+2 p_{z}^2}
{[\left(p_{0}-\frac{\omega }{2}\right)^2-\vec{p}^2-M^2] [\left(p_{0}+q_{0}+\frac{\omega }{2}\right)^2-\vec{p}^2-M^2]}\right\},
\end{aligned}
\eeq

\eq
\begin{aligned}
\Pi^{33}(q_{0})&=
-N_{f}N_{c}\int \frac{d^{4} p}{(2\pi)^{4}} \\
&\times
\left\{\frac{2 M^2-2 \left(p_{0}-\frac{\omega}{2}\right) \left(p_{0}+q_{0}-\frac{\omega}{2}\right)+2 p_{x}^2+2 p_{y}^2-2 p_{z}^2}
{[\left(p_{0}-\frac{\omega }{2}\right)^2-\vec{p}^2-M^2] [\left(p_{0}+q_{0}-\frac{\omega }{2}\right)^2-\vec{p}^2-M^2]}
+\frac{2 M^2-2 \left(p_{0}+\frac{\omega }{2}\right) \left(p_{0}+q_{0}+\frac{\omega }{2}\right)+2 p_{x}^2+2 p_{y}^2-2 p_{z}^2}
{[\left(p_{0}+\frac{\omega }{2}\right)^2-\vec{p}^2-M^2] [\left(p_{0}+q_{0}+\frac{\omega }{2}\right)^2-\vec{p}^2-M^2]}\right\}.
\end{aligned}
\eeq
We rewrite the relation in Eq.(\ref{coeff}):
\eq
\begin{aligned}
A_{1}^{2}&=-(\Pi_{11} - i \Pi_{12}),(s_{z}=-1 \text{ for } \rho \text{ meson }),\\
A_{2}^{2}&=-\Pi_{11} - i \Pi_{12},(s_{z}=+1\text{ for } \rho \text{ meson }),\\
A_{3}^{2}&=-\Pi_{33},(s_{z}=0 \text{ for } \rho \text{ meson }).
\end{aligned}
\eeq
The explicit form of coefficients can be given by:
\eq
\begin{aligned}
A^{2}_{1}(q_{0})=
2N_{f}N_{c}\int \frac{d^{4} p}{(2\pi)^{4}}
\frac{2 M^2-2 \left(p_{0}+\frac{\omega}{2}\right) \left(p_{0}+q_{0}-\frac{\omega}{2}\right)-2 p_{x}^2+2 p_{y}^2+2 p_{z}^2}
{\left[\left(p_{0}+\frac{\omega }{2}\right)^2-\vec{p}^2-M^2\right] \left[\left(p_{0}+q_{0}-\frac{\omega }{2}\right)^2-\vec{p}^2-M^2\right]},
\end{aligned}
\eeq

\eq
\begin{aligned}
A^{2}_{2}(q_{0})=
2N_{f}N_{c}\int \frac{d^{4} p}{(2\pi)^{4}}
\frac{2 M^2-2 \left(p_{0}-\frac{\omega }{2}\right) \left(p_{0}+q_{0}+\frac{\omega }{2}\right)-2 p_{x}^2+2 p_{y}^2+2 p_{z}^2}
{\left[\left(p_{0}-\frac{\omega }{2}\right)^2-\vec{p}^2-M^2\right] \left[\left(p_{0}+q_{0}+\frac{\omega }{2}\right)^2-\vec{p}^2-M^2\right]},
\end{aligned}
\eeq

\eq
\begin{aligned}
A^{2}_{3}(q_{0})=
N_{f}N_{c}\int \frac{d^{4} p}{(2\pi)^{4}}
&\left\{\frac{2 M^2-2 \left(p_{0}-\frac{\omega}{2}\right) \left(p_{0}+q_{0}-\frac{\omega}{2}\right)+2 p_{x}^2+2 p_{y}^2-2 p_{z}^2}
{\left[\left(p_{0}-\frac{\omega }{2}\right)^2-\vec{p}^2-M^2\right] \left[\left(p_{0}+q_{0}-\frac{\omega }{2}\right)^2-\vec{p}^2-M^2\right]}\right.\\
&+\left.\frac{2 M^2-2 \left(p_{0}+\frac{\omega }{2}\right) \left(p_{0}+q_{0}+\frac{\omega }{2}\right)+2 p_{x}^2+2 p_{y}^2-2 p_{z}^2}
{\left[\left(p_{0}+\frac{\omega }{2}\right)^2-\vec{p}^2-M^2\right] \left[\left(p_{0}+q_{0}+\frac{\omega }{2}\right)^2-\vec{p}^2-M^2\right]}\right\}.
\end{aligned}
\eeq
Now, it is obvious that
\eq
\frac{1}{2}A_{1}^{2}(m_{\rho}+\omega)+\frac{1}{2}A_{2}^{2}(m_{\rho}-\omega)
=A_{3}^{2}(m_{\rho}).
\eeq


\begin{thebibliography}{100}



\bibitem{Kharzeev:2007jp}
D.~E.~Kharzeev, L.~D.~McLerran and H.~J.~Warringa,
%``The Effects of topological charge change in heavy ion collisions: 'Event by event P and CP violation',''
Nucl. Phys. A \textbf{803}, 227 (2008)
doi:10.1016/j.nuclphysa.2008.02.298
[arXiv:0711.0950 [hep-ph]].
%1348 citations counted in INSPIRE as of 28 Jul 2020

\bibitem{Becattini:2007sr}
F.~Becattini, F.~Piccinini and J.~Rizzo,
%``Angular momentum conservation in heavy ion collisions at very high energy,''
Phys. Rev. C \textbf{77}, 024906 (2008)
doi:10.1103/PhysRevC.77.024906
[arXiv:0711.1253 [nucl-th]].
%169 citations counted in INSPIRE as of 28 Jul 2020

%\cite{Jiang:2016woz}
\bibitem{Jiang:2016woz}
Y.~Jiang, Z.~W.~Lin and J.~Liao,
%``Rotating quark-gluon plasma in relativistic heavy ion collisions,''
Phys. Rev. C \textbf{94}, no.4, 044910 (2016)
[erratum: Phys. Rev. C \textbf{95}, no.4, 049904 (2017)]
doi:10.1103/PhysRevC.94.044910
[arXiv:1602.06580 [hep-ph]].
%130 citations counted in INSPIRE as of 17 Nov 2020

\bibitem{Kharzeev:2007tn}
  D.~Kharzeev and A.~Zhitnitsky,
  %``Charge separation induced by P-odd bubbles in QCD matter,''
  Nucl.\ Phys.\  A {\bf 797}, 67 (2007).
%  [arXiv:0706.1026 [hep-ph]].
  %%CITATION = NUPHA,A797,67;%%

\bibitem{Son:2009tf}
  D.~T.~Son and P.~Surowka,
  %``Hydrodynamics with Triangle Anomalies,''
  Phys.\ Rev.\ Lett.\  {\bf 103}, 191601 (2009).
%  [arXiv:0906.5044 [hep-th]].
  %%CITATION = ARXIV:0906.5044;%%
  %182 citations counted in INSPIRE as of 04 Jan 2014

  %\cite{Kharzeev:2010gr}
\bibitem{Kharzeev:2010gr}
  D.~E.~Kharzeev and D.~T.~Son,
  %``Testing the chiral magnetic and chiral vortical effects in heavy ion collisions,''
  Phys.\ Rev.\ Lett.\  {\bf 106}, 062301 (2011).
  
%\cite{STAR:2017ckg}
\bibitem{STAR:2017ckg}
L.~Adamczyk \textit{et al.} [STAR],
%``Global $\Lambda$ hyperon polarization in nuclear collisions: evidence for the most vortical fluid,''
Nature \textbf{548}, 62-65 (2017)
doi:10.1038/nature23004
[arXiv:1701.06657 [nucl-ex]].
%300 citations counted in INSPIRE as of 17 Nov 2020

%\cite{Acharya:2019vpe}
\bibitem{Acharya:2019vpe}
S.~Acharya \textit{et al.} [ALICE],
%``Evidence of Spin-Orbital Angular Momentum Interactions in Relativistic Heavy-Ion Collisions,''
Phys. Rev. Lett. \textbf{125}, no.1, 012301 (2020)
doi:10.1103/PhysRevLett.125.012301
[arXiv:1910.14408 [nucl-ex]].
%18 citations counted in INSPIRE as of 17 Nov 2020

%\cite{Klevansky:1992qe}
\bibitem{Klevansky:1992qe}
S.~P.~Klevansky,
%``The Nambu-Jona-Lasinio model of quantum chromodynamics,''
Rev. Mod. Phys. \textbf{64}, 649-708 (1992)
doi:10.1103/RevModPhys.64.649
%1608 citations counted in INSPIRE as of 12 Aug 2020

%\cite{Hidaka:2012mz}
\bibitem{Hidaka:2012mz}
Y.~Hidaka and A.~Yamamoto,
%``Charged vector mesons in a strong magnetic field,''
Phys. Rev. D \textbf{87}, no.9, 094502 (2013)
doi:10.1103/PhysRevD.87.094502
[arXiv:1209.0007 [hep-ph]].
%95 citations counted in INSPIRE as of 12 Aug 2020

%\cite{Liu:2014uwa}
\bibitem{Liu:2014uwa}
H.~Liu, L.~Yu and M.~Huang,
%``Charged and neutral vector $\rho$ mesons in a magnetic field,''
Phys. Rev. D \textbf{91}, no.1, 014017 (2015)
doi:10.1103/PhysRevD.91.014017
[arXiv:1408.1318 [hep-ph]].
%46 citations counted in INSPIRE as of 12 Aug 2020

%\cite{Liu:2018zag}
\bibitem{Liu:2018zag}
H.~Liu, X.~Wang, L.~Yu and M.~Huang,
%``Neutral and charged scalar mesons, pseudoscalar mesons, and diquarks in magnetic fields,''
Phys. Rev. D \textbf{97}, no.7, 076008 (2018)
doi:10.1103/PhysRevD.97.076008
[arXiv:1801.02174 [hep-ph]].
%20 citations counted in INSPIRE as of 12 Aug 2020

%\cite{Yamamoto:2013zwa}
\bibitem{Yamamoto:2013zwa}
A.~Yamamoto and Y.~Hirono,
%``Lattice QCD in rotating frames,''
Phys. Rev. Lett. \textbf{111}, 081601 (2013)
doi:10.1103/PhysRevLett.111.081601
[arXiv:1303.6292 [hep-lat]].
%31 citations counted in INSPIRE as of 07 Aug 2020

\bibitem{Matsuo:2015}
Matsuo.Mamoru,Ieda.Junichi,Maekawa.Sadamichi
%TITLE=Mechanical generation of spin current
Frontiers in Physics,{\bf 54}, no.3, 00054 (2015)
%VOLUME=3YEAR=2015PAGES=54
%URL=https://www.frontiersin.org/article/10.3389/fphy.2015.00054
doi:10.3389/fphy.2015.00054

%\cite{Jiang:2016wvv}
\bibitem{Jiang:2016wvv}
Y.~Jiang and J.~Liao,
%``Pairing Phase Transitions of Matter under Rotation,''
Phys. Rev. Lett. \textbf{117}, no.19, 192302 (2016)
doi:10.1103/PhysRevLett.117.192302
[arXiv:1606.03808 [hep-ph]].
%42 citations counted in INSPIRE as of 12 Aug 2020

%\cite{Wang:2018sur}
\bibitem{Wang:2018sur}
X.~Wang, M.~Wei, Z.~Li and M.~Huang,
%``Quark matter under rotation in the NJL model with vector interaction,''
Phys. Rev. D \textbf{99}, no.1, 016018 (2019)
doi:10.1103/PhysRevD.99.016018
[arXiv:1808.01931 [hep-ph]].
%9 citations counted in INSPIRE as of 09 Aug 2020

%\cite{Zhang:2018ome}
\bibitem{Zhang:2018ome}
H.~Zhang, D.~Hou and J.~Liao,
%``Mesonic Condensation in Isospin Matter under Rotation,''
[arXiv:1812.11787 [hep-ph]].
%8 citations counted in INSPIRE as of 12 Aug 2020

%\cite{Chen:2015hfc}
\bibitem{Chen:2015hfc}
H.~L.~Chen, K.~Fukushima, X.~G.~Huang and K.~Mameda,
%``Analogy between rotation and density for Dirac fermions in a magnetic field,''
Phys. Rev. D \textbf{93}, no.10, 104052 (2016)
doi:10.1103/PhysRevD.93.104052
[arXiv:1512.08974 [hep-ph]].
%44 citations counted in INSPIRE as of 12 Aug 2020

%\cite{Matsuo:2012wv}
\bibitem{Matsuo:2012wv}
M.~Matsuo, J.~Ieda and S.~Maekawa,
%``Renormalization of spin-rotation coupling,''
Phys. Rev. B \textbf{87}, 115301 (2013)
doi:10.1103/PhysRevB.87.115301
[arXiv:1211.0127 [cond-mat.mes-hall]].
%5 citations counted in INSPIRE as of 08 Aug 2020

%\cite{McInnes:2016dwk}
\bibitem{McInnes:2016dwk}
B.~McInnes,
%``A rotation/magnetism analogy for the quark篓Cgluon plasma,''
Nucl. Phys. B \textbf{911}, 173-190 (2016)
doi:10.1016/j.nuclphysb.2016.08.001
[arXiv:1604.03669 [hep-th]].
%14 citations counted in INSPIRE as of 12 Aug 2020

\bibitem{Bernard:1988db}
  V.~Bernard and U.~G.~Meissner,
  %``Properties of Vector and Axial Vector Mesons from a Generalized Nambu-Jona-Lasinio Model,''
  Nucl.\ Phys.\ A {\bf 489}, 647 (1988).
  doi:10.1016/0375-9474(88)90114-5
  %%CITATION = doi:10.1016/0375-9474(88)90114-5;%%
  %197 citations counted in INSPIRE as of 12 Aug 2018

\bibitem{Buballa:2003qv}
  M.~Buballa,
  %``NJL model analysis of quark matter at large density,''
  Phys.\ Rept.\  {\bf 407}, 205 (2005)
  doi:10.1016/j.physrep.2004.11.004
  [hep-ph/0402234].
  %%CITATION = doi:10.1016/j.physrep.2004.11.004;%%
  %947 citations counted in INSPIRE as of 30 Jul 2018

\bibitem{He:1997gn}
	Y.~B.~He, J.~Hufner, S.~P.~Klevansky and P.~Rehberg,
	%``Pi pi scattering in the rho meson channel at finite temperature,''
	Nucl.\ Phys.\ A {\bf 630}, 719 (1998)
	[nucl-th/9712051].
	%%CITATION = NUCL-TH/9712051;%%
	%18 citations counted in INSPIRE as of 23 Jun 2014

\bibitem{Rehberg:1995nr}
	P.~Rehberg and S.~P.~Klevansky,
	%``One loop integrals at finite temperature and density,''
	Annals Phys.\  {\bf 252}, 422 (1996)
	[hep-ph/9510221].
	%%CITATION = HEP-PH/9510221;%%
	%9 citations counted in INSPIRE as of 25 Jun 2014

\bibitem{Kapusta}
  J.I. Kapusta, {\it Finite Temperature Field Theory},
  Cambridge University Press, Cambridge (1989).

%\cite{Frasca:2011zn}
\bibitem{Frasca:2011zn}
M.~Frasca and M.~Ruggieri,
%``Magnetic Susceptibility of the Quark Condensate and Polarization from Chiral Models,''
Phys. Rev. D \textbf{83}, 094024 (2011)
doi:10.1103/PhysRevD.83.094024
[arXiv:1103.1194 [hep-ph]].
%69 citations counted in INSPIRE as of 03 Nov 2020

\bibitem{Tatsumi-spinpolarization}
%\bibitem{Tatsumi:1999ab} 
T.~Tatsumi,
%``Ferromagnetism of quark liquid,''
Phys.\ Lett.\ B {\bf 489}, 280 (2000)
doi:10.1016/S0370-2693(00)00927-8
[hep-ph/9910470].
%%CITATION = doi:10.1016/S0370-2693(00)00927-8;%%
%122 citations counted in INSPIRE as of 28 Jun 2020
%\cite{Maruyama:2000cw}
%\bibitem{Maruyama:2000cw} 
T.~Maruyama and T.~Tatsumi,
%``Ferromagnetism of nuclear matter in the relativistic approach,''
Nucl.\ Phys.\ A {\bf 693}, 710 (2001)
doi:10.1016/S0375-9474(01)00811-9
[nucl-th/0010018].
%%CITATION = doi:10.1016/S0375-9474(01)00811-9;%%
%42 citations counted in INSPIRE as of 28 Jun 2020
%\cite{Nakano:2003rd}
%\bibitem{Nakano:2003rd} 
E.~Nakano, T.~Maruyama and T.~Tatsumi,
%``Spin polarization and color superconductivity in quark matter,''
Phys.\ Rev.\ D {\bf 68}, 105001 (2003)
doi:10.1103/PhysRevD.68.105001
[hep-ph/0304223].
%%CITATION = doi:10.1103/PhysRevD.68.105001;%%
%61 citations counted in INSPIRE as of 28 Jun 2020
%\cite{Tatsumi:2003bk}
%\bibitem{Tatsumi:2003bk} 
T.~Tatsumi, T.~Maruyama and E.~Nakano,
%``Ferromagnetism and superconductivity in quark matter: Color magnetic superconductivity,''
Prog.\ Theor.\ Phys.\ Suppl.\  {\bf 153}, 190 (2004)
doi:10.1143/PTPS.153.190
[hep-ph/0312347].
%%CITATION = doi:10.1143/PTPS.153.190;%%
%36 citations counted in INSPIRE as of 28 Jun 2020T
\end{thebibliography}
\end{document}